\documentclass[journal=jctcce,manuscript=article,etalmode=truncate,maxauthors=10,layout=twocolumn,9pt]{achemso}   

\setkeys{acs}{etalmode=truncate,maxauthors=10}  

\usepackage{chemformula} 
\usepackage[T1]{fontenc} 

\usepackage{amsmath}
\usepackage{graphicx}
\usepackage{wrapfig}
\usepackage{xcolor}  

\usepackage{balance}   

\usepackage[colorlinks=true, allcolors=blue]{hyperref}

\usepackage{physics}
\usepackage[makeroom]{cancel} 

\usepackage{algorithm}

\usepackage{algpseudocode}  

\author{Jacek Jakowski}
\affiliation{Center For Nanophase Materials Sciences,  Oak Ridge National Laboratory, Oak Ridge, TN 37831, USA}
\alsoaffiliation{Computational  Sciences and Engineering Division, Oak Ridge National Laboratory, Oak Ridge, TN 37831, USA}
\email{jakowskij@ornl.gov}
\phone{+1 865 574 6174}

\author{Wenchang Lu }
\affiliation{Department of Physics, North Carolina State University, Raleigh, NC 27695-8202, USA}

\author{Emil Briggs}
\affiliation{Department of Physics, North Carolina State University, Raleigh, NC 27695-8202, USA}

\author{David Lingerfelt}
\affiliation{Center For Nanophase Materials Sciences,  Oak Ridge National Laboratory, Oak Ridge, TN 37831, USA}

\author{Bobby G. Sumpter}
\affiliation{Center For Nanophase Materials Sciences,  Oak Ridge National Laboratory, Oak Ridge, TN 37831, USA}

\author{Panchapakesan Ganesh}
\affiliation{Center For Nanophase Materials Sciences,  Oak Ridge National Laboratory, Oak Ridge, TN 37831, USA}

\author{Jerzy Bernholc}
\affiliation{Department of Physics, North Carolina State University, Raleigh, NC 27695-8202, USA}

\title{\large 
Simulation of  24,000 Electrons Dynamics: Real-Time Time-Dependent Density Functional Theory (TDDFT) with the Real-Space Multigrids  (RMG)\footnote{
NOTICE OF COPYRIGHT: This manuscript has been authored by UT-Battelle, LLC under Contract No. DE-AC05-00OR22725 with the U.S. Department of Energy. The United States Government retains and the publisher, by accepting the article for publication, acknowledges that the United States Government retains a non-exclusive, paid-up, irrevocable, worldwide license to publish or reproduce the published form of this manuscript, or allow others to do so, for United States Government purposes. The Department of Energy will provide public access to these results of federally sponsored research in accordance with the DOE Public Access Plan (http://energy.gov/downloads/doe-public-access-plan).
 } }

\abbreviations{DFT, TD-DFT}

\begin{document}

\footnotesize

\begin{abstract}  
We present the theory, implementation, and benchmarking of a real-time time-dependent density functional theory (RT-TDDFT) module within the RMG code, designed to simulate the electronic response of molecular systems to external perturbations. Our method offers insights into non-equilibrium dynamics and excited states across a diverse range of systems, from small organic molecules to large metallic nanoparticles. Benchmarking results demonstrate excellent agreement with established TDDFT implementations and showcase the superior stability of our time-integration algorithm, enabling long-term simulations with minimal energy drift. The scalability and efficiency of RMG on massively parallel architectures allow for simulations of complex systems, such as plasmonic nanoparticles with thousands of atoms. Future extensions, including nuclear and spin dynamics, will broaden the applicability of this RT-TDDFT implementation, providing a powerful toolset for studies of photoactive materials, nanoscale devices, and other systems where real-time electronic dynamics is essential.

\end{abstract}


\section{Introduction}

The application of time-dependent electronic structure methods allows for the modeling of molecular response to external perturbations, such as external fields or charged particle beams, providing a wealth of information about physicochemical processes, including non-equilibrium dynamics or electronically excited states \cite{tddft-fundamentals,chem-rev-Li}. In-depth discussions on the theoretical foundations and practical applications of TDDFT for linear response (LR) in the frequency domain and real-time (RT) in the time domain can be found in Ref. [\citenum{tddft-fundamentals}].

Historically,  one of the first notable implementations of real-time (RT)  time-dependent density functional theory (TDDFT) has  been reported by  Bertsch and Yabana \cite{bertsch-96}. It   was based  on a real-space grid  and the local density approximation (LDA)  and applied  to benzene.
In the following decades, RT-TDDFT 
took a central  role  and became the method of choice for applications  in spectroscopy \cite{chem-rev-Li}, materials science, solar energy conversion and optoelectronics\cite{review-photovoltaics}, photo- and femto-chemistry, nanoscale plasmonics \cite{LSPR-review}, and photocatalysis \cite{chem-rev-Li,CO2-AuNP}.  
Recent examples include 
(1) electronic circular dichroism, optical UV/VIS and near edge x-ray  absorption, Raman scattering spectroscopy\cite{lopata-2011,raman-govind-2013,Yabana-dichrosim-2011,Rubio-angular-2017,xray-herbert-2023}, 
(2) nonlinear response of molecular systems in strong fields \cite{Mukamel-2018,non-linear-Gross-2015,Li-hyperpolarizabilities-2013}, 
(3) charge transfer, photovoltaics,  and solar energy conversion, redox reaction \cite{review-photovoltaics,chem-rev-Li,charge-transfer-isborn-2023,lvn-ct-2012}, 
(4)  real-time electron transport and molecular devices  \cite{rt-transport-Kummel-2016,rt-transport-hod-jpca-2016,rt-transport-hod-jpcc-2016},
(5) ion and electron beam interactions with materials \cite{AccMatRes-2024,theo-comp-chem-DBL-2022,theo-comp-chem-JJ-2022,JCTC-2020,tao-2021}, 
(6) stopping power \cite{baczewski-23,schleife,Ievlev}, 
(7) nanoplasmonics and plasmonic catalysis \cite{wong-plasmonic-antennas-2017,wong-plasmonics-waveguide-2018,plasmonic-catalysis-montemore,tao-2021}, 
(8) spin  and  magnetization dynamics \cite{spin-rt-tddft-Li-2014,spin-rt-tddft-Li-2016,x2c-Ruud-2016,LvN-spin-Hod-Scuseria-2015}.

The real-time TDDFT methodology has been implemented in many electronic structure  codes,  ranging from  quantum chemistry  codes  utilizing gaussian (\texttt{NWchem, Gaussian, Q-Chem}) and numerical (\texttt{Siesta, DFTB+}) atom-centered basis sets, \cite{lopata-2011,rt-dftb-Wong,rt-tddft-qchem,siesta-2007,DFTB+} materials science periodic unit cell codes based on plane waves (\texttt{Qbox/Qball, cp2k, Exciting}) and real space grids (\texttt{Octopus}), \cite{octopus,qball-2017,qball-II-2021,cp2k-2020}  to finite element codes (\texttt{DFT-FE}) \cite{gavini-rt-tddft}.   
Several extensions of real-time electronic structure methods go  beyond DFT,  including GW  methods  \cite{rt-GW-jctc-2019,rt-GW-rabani-2014}, BSE \cite{rt-BSE-2011}, configuration interaction, coupled clusters\cite{Gordon-2012,tdci-saalfrank-2007,tdci-Li-2018,rt-cc-dePrince-2016},  relativistic methods \cite{x2c-Ruud-2016,spin-rt-tddft-Li-2016}, and  coupled  electron  nuclei dynamics  \cite{Thruhlar-2018,ExactFactorization-2012,namd-dftb,ehrenfest-gaussian-2007,lvnmd,lvnmd-c2,rt-tddft-md-siesta-2016,pyxaid-2013,qball-II-2021}.

Compared to frequency-domain linear response (LR) TDDFT, real-time (RT) TDDFT offers several advantages. RT-TDDFT is an initial value boundary problem. Depending on the choice of the initial electronic state and boundary conditions, it is possible to model non-equilibrium charge transfer or electron transport \cite{chem-rev-Li}. RT-TDDFT allows simulation of electronic response to perturbations of arbitrary shape and duration \cite{chem-rev-Li}. Its computational scaling and resource requirements are similar to conventional DFT methods, enabling simulations of large molecular systems not accessible to LR-TDDFT, such as the 20,000-electron system simulations demonstrated with the plane wave Qbox/Qball code \cite{qbox-2008,qball-2017,qball-II-2021}. However, RT-TDDFT can suffer from peak shifts and inaccurate charge transfer descriptions in non-equilibrium electron dynamics, attributed to the adiabatic DFT approximation and its instantaneous time-dependent densities \cite{peak-shift-isborn-2015,peak-shift-maitra-2016,ct-maitra-2013,ct-maitra-2014}. Similarly, the coupled electron-nuclei molecular dynamics utilizing instantaneous RT-TDDFT-based electronic structure (so-called Ehrenfest dynamics) follow a "mean field" path and struggle to reproduce the branching into individual reaction pathways expected in the long-time stochastic regime \cite{Thruhlar-2018}.

In general, the choices of representation or basis set used to describe the electronic structure in DFT and TD-DFT offer different advantages and disadvantages. The main approaches include: (1) Localized atomic-like orbitals, 
 which are highly efficient for systems with localized electronic levels, such as molecular systems, but can show unphysical intruder peaks  above the ionization threshold,  which are artifacts of basis set incompleteness  and correspond to transitions to short-lived (unbound) scattering states in the  continuum.\cite{qchem5,chem-rev-Li,lopata-2011,DFTB+}  
(2) Plane waves, which are advantageous for periodic systems due to their natural handling of periodic boundary conditions and uniform accuracy across the unit cell, but  are inefficient in dealing with localized and charged systems,  and require pseudopotentials and high kinetic energy cutoffs to achieve convergence \cite{octopus,qball-2017,qball-II-2021,QE-TDDFT2}. 
 
Uniformly-spaced grids, which we utilize in this work, are dual to plane waves and thus provide a spatially unbiased basis set that is independent of atomic positions and offers a single accuracy criterion, the grid spacing, equivalent to the plane-wave kinetic energy cutoff used in plane-wave methods \cite{rmg-dft-2024}. However, plane waves offer a natural and very accurate handling of the kinetic energy, while real-space grids require adaptive discretization of the kinetic energy operator to reach the same accuracy as plane waves.\cite{rmg-dft-2024} Real-space grid-based methods are also flexible in handling boundary conditions.  Both periodic and non-periodic systems can be easily handled, making them versatile for a wide range of systems, from crystals and bulk materials to molecules and surfaces. 
They also allow for potential simulations of processes involving a free electron ejected by light or a beam particle to a continuum state, electron tunneling from a surface electrode, or the chemistry of a solvated electron.

In contrast to the plane-wave approach, real-space methods
are inherently local and can be efficiently implemented on massively parallel computer architectures,
with each processor assigned to a different region of space.
The \texttt{RMG} code used in this work has been implemented on many highly parallel systems, including exascale supercomputers (Frontier at ORNL and Aurora at ANL), enabling simulations of very large systems consisting of thousands of atoms.

The rest of this paper is organized as follows.  
The next section \ref{sec:methodology} discusses the  computational  methodology, starting  
with  a brief overview of the RMG method and code, followed  by theoretical  foundations 
of density matrix propagation based  on commutator expansion. The derivations of the commutator expansion  are general 
and can be used beyond the scope of this work.
In section \ref{sec:implementation}, we describe the algorithm used here, its implementation, and code optimization for parallel execution.
Section \ref{sec:results} presents  applications to simulations of optical response of  small  and medium  size  molecules 
(benzene, C$_{60}$ fullerene and chlorophylls) and compares the results with other electronic structure codes. The last part of this section  highlights  the scaling and performance of  the method in  applications
to the plasmonic response of large  silver nanoparticles.
Such systems are  of primary importance  for  photo- and electro-catalysis on plasmonic  nanoparticles, 
and  the development of novel sensors based on  localized surface plasmon resonance \cite{CO2-AuNP,LSPR-review}.
The paper concludes with  a Summary and Outlook section.

\section{Methodology}
\label{sec:methodology}
Here  we discuss  the  computational and  theoretical background behind the current  implementation 
of  real-time TDDFT  in the RMG  suite of codes,
starting with a brief overview of RMG capabilities. 
Next, we describe the theoretical foundations  behind  our computational approach
in the context of  DFT implementation in RMG, including  the Magnus expansion and  density matrix propagation.
The commutator  expansion that we present allows to effectively propagate  density matrices
 at the   accuracy of exact diagonalization propagation.
The  derivation of commutator expansion used for  density matrix propagation is general  and can  be used to formulate
other efficient computation  schemes  beyond the current work, 
to address  such topics as electron dynamics governed by  non-Hermitian Hamiltonians, or combined electron-ion dynamics.


\subsection{ Overview of RMG }

%
\texttt{RMG} (Real-space MultiGrid) is a Density Functional Theory (DFT)-based electronic structure code \cite{briggs1996,rmg-dft-2024} designed for scalability and high performance on massively parallel supercomputers from its inception. The DFT equations \cite{kohn-1964,Kohn-1865} are discretized on real-space grids and adopt the following form: 
\begin{align} 
\vb H[\Psi_n] = \frac{1}{2}\Delta [\Psi_n] +V_{eff}\Psi_n = \varepsilon_n \vb S [\Psi_n] 
\label{eq:DFT}
\end{align}
where $\vb H$ denotes the Hamiltonian consisting of the kinetic energy and the potential terms, with $\Psi_n$  representing the $n^\text{th}$ single-electron Kohn-Sham orbital ($n=1,\ldots,N)$ and $\vb S$ being the overlap matrix. A precise, multi-order adaptive discretization is used for the Laplacian, resulting in the same average accuracy as plane-wave codes on the well-known Delta test for 71 elements in the periodic table \cite{rmg-dft-2024,DFT-reproducibility}. 

In the simplest one-dimensional (1D) grid case, the Laplacian can be represented as a banded symmetric matrix. Its effect on a vector $\Psi(x)$, discretized on a uniform grid with spacing $h$, is approximated using the central finite difference formula with $2n+1$ grid points:
\begin{align} 
\Psi^{''}(x_0) =& a_0 \cdot \Psi(x_0) + \sum_{i=1}^{n} a_i \cdot \big(\Psi(x_0+ ih) + \Psi(x_0-ih) \big) \nonumber \\ 
     & + R_n(h^{2n})
\end{align}
where  $R_n(h^{2n})$ is the truncation error and the coefficients  $\{a_0, a_1, \ldots a_n \}$ 
are optimized to make the expression exact for polynomials of degree up to
$2n$. In the adaptive discretization used in RMG, a mixture of two Laplacian discretizations, typically of orders 8 and 6, is employed, according to the formula\cite{rmg-dft-2024}
\begin{align} 
\Psi^{''}_{(new)}(x_0) =& (1+M)\Psi^{''}_{(n)}(x_0) -M\Psi^{''}_{(n-1)}(x_0)
\end{align}
where $n=4$ is used. The value of the non-negative mixing parameter $M$ is optimized by minimizing the error in the Laplacian for a given combination of species using atomic wavefunctions at the beginning of the calculations. The optimization procedure, which is very fast, has been extensively tested and described in the original publication.\cite{rmg-dft-2024} 

The expectation value of an arbitrary operator $A$ is computed using the standard expression:
\begin{align}  
\bra{\Psi} A \ket{\Psi} = h \sum_{i,j}     \Psi(x_i) \cdot A(x_i,x_j) \cdot \Psi(x_j)
\end{align}
The potential term $V_{eff}$ in Eq. \ref{eq:DFT} contains contributions from the classical electron-ion attraction (with ions represented by nonlocal or semi-local pseudopotentials), the electron-electron electrostatic repulsion, and an exchange-correlation term, which may include nonlocal contributions (e.g., Hartree-Fock, hybrid, and dispersion functionals). The $\vb S$ matrix is non-diagonal when ultrasoft pseudopotentials are used \cite{Vanderbilt-1990}.

Multigrid preconditioning is used to accelerate convergence by employing a sequence of grids of varying resolutions. The solution is obtained on a grid fine enough to accurately represent both the potentials and the wavefunctions. It is well known that iterations on a given grid level reduce errors with wavelengths comparable to the grid spacing but are less effective for longer wavelengths \cite{multigrids-Brandt-2011,multigrids-Briggs-2000}. Lower-frequency components are handled on auxiliary grids with progressively larger spacings, where remaining errors appear as high-frequency components. This approach ensures rapid convergence. Extensive comparisons show that \texttt{RMG} agrees extremely well \cite{rmg-dft-2024} with the plane-wave \texttt{Quantum Espresso} code \cite{QE-2017}. The reasons for this close agreement are that plane waves and uniformly spaced real-space grids are duals, and \texttt{RMG} uses very accurate discretizations.

Similarly to plane-wave approaches, pseudopotentials are employed in \texttt{RMG} to represent ionic cores, and both norm-conserving \cite{Hamann,BHS,KleinmanBylander} and ultrasoft \cite{Vanderbilt} pseudopotentials are supported. Achieving accuracy equivalent to plane-wave codes at grid resolutions similar to plane-wave kinetic energy cutoffs requires adaptive discretization of the kinetic energy operator \cite{rmg-dft-2024}. To facilitate large-scale calculations, two approaches are implemented for computing the electronic structure: (i) delocalized orbitals, and (ii) localized orbitals. In the former approach, the orbitals are delocalized across the entire simulation cell, closely analogous to plane-wave calculations, and achieve equivalent accuracy. The latter approach relies on the variational construction of an optimized minimal basis set of atom-centered orbitals localized within a predetermined cutoff radius, reducing overall computational cost and scaling \cite{fatt-bern,fattebert}, while reproducing the DFT total energy of the corresponding delocalized-orbitals calculations within 1 $mHa$ accuracy.

The \texttt{RMG} code suite consists of several modules. The  distribution (\url{https://github.com/RMGDFT}) contains the code needed to perform DFT calculations for a large number of atoms, and molecular dynamics with DFT forces \cite{RMG-DFT-github}. An additional non-Equilibrium Green's Function (NEGF) module allows for quantum transport calculations\cite{Lu_PRL-2005} for nanoscale devices using an optimized localized basis. 
RMG has been tested on a large number of systems, including supercells of bulk materials containing various impurities, surfaces,  interfaces, transition metal oxides, and 2D materials. Its results agree very well with plane-wave calculations and experiment \cite{briggs1996,rmg-dft-2024}. 

\texttt{RMG} runs on Linux, Windows and OS X. It scales from laptops, desktops and clusters to the largest supercomputers, utilizing all CPU cores and GPUs of each cluster or supercomputer node. It also provides large-scale DFT input to the Quantum Monte Carlo \texttt{QMCPACK} codes,\cite{qmcpack1,qmcpack2} and was part of the QMCPACK Exascale Computing Program project.\cite{exascale} 
The web based  interface has been implemented and is available online (\url{https://tinyurl.com/rmgdft-web}) allowing users to easily generate RMG text input files and automatically populate all control parameters \cite{RMG-DFT-gui}.
The  RMG web interface accept atomic coordinate files in several formats. 
\texttt{RMG} documentation is available at \url{https://github.com/RMGDFT/rmgdft/wiki}.

\subsection{Theory } %
\label{sec:theory}

The real-time Time-Dependent Density Functional Theory (RT-TDDFT) problem, which is central to this paper, can be viewed as a special case within the broader mathematical framework of strongly continuous C$_0$ semigroups \cite{semigroups}. While the time-dependent Schr\"odinger equation, which governs RT-TDDFT, describes unitary propagation in quantum mechanics, the C$_0$  semigroup framework also applies to non-unitary evolution processes governed by other types of partial differential equations.

For example, C$_0$  semigroups can describe the solutions to hyperbolic equations, such as the wave equation, which involves reversible wave propagation, as well as parabolic equations like the diffusion and heat equations, which describe irreversible processes. In these cases, the non-unitary nature reflects the dissipative or smoothing effects inherent to these systems.

The C$_0$ semigroup is a one-parameter family of operators that depends on a time parameter, and the semigroup property (lack of inverse elements) reflects the irreversibility of certain processes, such as diffusion and heat conduction described by parabolic equations. Importantly, the solutions within this framework are often expressed through the direct exponentiation of the semigroup generator, providing a fundamental connection between the mathematical theory and the physical evolution of the system. This connection will also be demonstrated in this manuscript.

\subsubsection{Magnus expansion} 

The exact solution  to
 the time-dependent Schrodinger equation
\begin{equation}
\imath \hbar \frac{\partial}{\partial t} | \Psi(t,x) \rangle  =  \hat H(x,t)  |\Psi(t,x) \rangle,
\end{equation}
where $\Psi(t,x)$  is  the time-dependent wave function  
subject to the initial conditions $\Psi(0,x)=\Psi_0(x)$  
is given by 
\begin{align}
 \Psi(t,x)  =&  {\mathcal U}(t) \Psi(0,x),  
\end{align}
where  ${\mathcal U}(t) $ is  time evolution  operator equal  to
\begin{equation}
 {\mathcal U}(t)  = \exp( \Omega(t))
\label{eq:U_operator}
\end{equation}
with prescription for $\Omega(t)$ given by the Magnus expansion \cite{Magnus-1954,Oteo-2000,Casas-2007,lvnmd} 
The $\Omega(t)$ in the time-evolution operator is a time-ordered  Hamiltonian integral  that can 
formally be written as $\Omega(t)=-\frac{\imath}{\hbar} \int_0^t \mathcal H(\tau)d\tau$. 
Here, the symbol $\mathcal H(\tau)$ represents time-ordering of the Hamiltonian, $H(t)$,
 to distinguish it from the  usual integration.
Following the Magnus expansion,  $\Omega(t)$ can be calculated according to 
\begin{align} 
\Omega_{~}   = &  \Omega_1 +  \Omega_2 +  \Omega_3 \ldots  \label{eq:omega} \\
\Omega_1 = & \frac{-\imath}{\hbar} \int_0^t  H(\tau) d\tau_1  \label{eq:omega1}  \\
\Omega_2 = & \frac{1}{2}  \int_0^t d\tau_1 \int_0^{\tau_1} d\tau_2   
  \left[\frac{ H(\tau_1)}{\imath \hbar}, \frac{ H(\tau_2)}{\imath \hbar}\right] \label{eq:omega2} \\
 \nonumber \Omega_3 = &  \frac{1}{2}  \int_0^t d\tau_1 \int_0^{\tau_1} d\tau_2 \int_0^{\tau_2} d\tau_3 
  \Big(\Big[\frac{ H(\tau_1)}{\imath \hbar}, \big[ \frac{ H(\tau_2)}{\imath \hbar},\frac{ H(\tau_3)}{\imath \hbar}\big] \Big]    \\
 & + \Big[\frac{H(\tau_3)}{\imath \hbar}, \big[\frac{ H(\tau_2)}{\imath \hbar},\frac{ H(\tau_1)}{\imath \hbar}\big] \Big] \Big)    \label{eq:omega3}
\end{align} 
with the square bracket denoting the usual commutator $[A,B]=AB-BA$.
In all practical implementations, the propagation is performed using a small  time  step $\Delta t$  
so that the change of the Hamiltonian $H(t)$  within the time step is linear
$H(t) = H_0 + t \cdot H'$,  where  $H_0=H(0)$,  $H_1= H(\Delta t)$  and  $H' = \frac{H_1 -H_0}{\Delta t}$.
This allows for analytical  integration of the time-ordering commmutator expressions,
 $\Omega_k$ in the  Magnus expansion, with  
a first few  terms given by 
\begin{align} 
\label{eq:omega1a} \Omega_1 = & -\frac{\imath}{\hbar} \cdot \tfrac{1} {2} (H_0 +H_1) \Delta t   \\
\Omega_2 = & \frac{1}{12 \hbar^2} \Delta t^2 [H_0, H_1 ]  
\end{align}
where we assumed  that the matrix elements of  $H(t)$ change linearly with time from $H_0$ to $H_1$.
Here, $\tfrac{1} {2} (H_0 +H_1)$ in  $\Omega_1$ describes an average Hamiltonian between $t=0$ and $t=\Delta t$.

Inserting  expressions for $\Omega_1$, $\Omega_2\ldots$ into Eq. \ref{eq:U_operator}
provides  a  foundation  for formulations  of different symplectic and unitary propagation schemes,
 which are  particularly suitable for long-time quantum dynamics simulations.
For example, the exact  time-evolution operator  can be constructed in a spectral form through diagonalization
${\mathcal U}(t) = {\mathbf W} \cdot \exp(\omega)\cdot {\mathbf W}^\dag$, where $\omega$ is a diagonal matrix 
of eigenvalues of $\Omega$ matrix operator and $ {\mathbf W}$ are corresponding eigenvectors.
 Limiting $\Omega$ to the first term, $\Omega_1$, leads to a  mid-point propagation scheme. 
Next, Taylor expansion of $\exp(\Omega_1)$ truncated to the first term  (linear in time) leads to the Crank-Nicolson
propagation scheme. The Runge-Kutta propagation is not symplectic and as such is not suitable for long-time 
 simulations.

\subsubsection{Density matrix propagation  }

 In general, TDDFT equations can be solved using either the state vector approach or density matrix propagation, as the two approaches are equivalent in many cases. However, the density matrix approach offers several advantages: it naturally allows for the description of mixed thermal states and can be extended to incorporate interactions with the environment via density matrix embedding techniques or Lindblad-type master equations. This makes it particularly useful for exploring non-equilibrium processes and dissipation mechanisms. Furthermore, the density matrix formulation is numerically robust against noise accumulation during long-time dynamics and facilitates efficient parallelization expressed in terms of matrix-matrix operations.
 
Consider  the one-particle density matrix,  
$P(t)= \sum_i f_i | \Psi_i(t)\rangle\langle  \Psi_i(t) |$,  
where the summation runs over all states and $f_i$ are the corresponding  occupation numbers.
Its time evolution is governed by  the Liouville-von Neumann equation
\begin{equation}
\frac{\partial P(t)}{ \partial t} = -\frac{\imath}{\hbar} \left( H(t) P(t) - P(t) H(t) \right).
\label{eq:LvN} 
\end{equation}
The general expression for the  time evolution of density matrices 
 is given by
\begin{align}
\label{eq:UxP0xU}
P(t) =& {\mathcal U}(t)\cdot P(0)\cdot {\mathcal U}(t)^\dag \\ \nonumber
     =& \exp(\Omega) \cdot P(0)\cdot  \exp(\Omega^\dag)  
\end{align}
where ${\mathcal U}(t)$  can be directly obtained in a spectral form from  diagonalization of the $\Omega$ operator.

We will now show a derivation of an alternative scheme that avoids diagonalization.
We start by inserting an auxilary parameter $\lambda$ into the time evolution operator for  density matrix 
propagation, Eq. \ref{eq:UxP0xU},  
\begin{align} 
P(\lambda) =& \exp(\lambda \Omega) \cdot P(0)\cdot \exp(\lambda \Omega)^\dag \\ \nonumber 
  =&  \exp(\lambda \Omega) \cdot P(0)\cdot \exp(\lambda \Omega^\dag ),
\label{eq:P-propagation}
\end{align}
so that  for $\lambda=1$ the density matrix $P(\lambda)=P(t)$, whereas for  $\lambda=0$  it becomes
 an initial (unpropagated) density matrix $P(0)$.
Here $\Omega \equiv   \Omega(t)$  is  a time-ordered operator  described  by Eqs.  \ref{eq:omega}-\ref{eq:omega3}.
We will now find  $P(\lambda=1)$ from the  Taylor expansion of  $P(\lambda)$  around  $\lambda=0$
\begin{equation}
P(\lambda)  = \sum_{k=0}^{+\infty}  \frac{\lambda^k}{k!} 
  P^{(k)} (0). 
\label{eq:Taylor}
\end{equation}
Here,  $P^{(k)}(0) = \left.\frac{d^k}{d\lambda^k}  P(\lambda) \right|_{\lambda=0}$.
The  derivatives  $ P^{(k)} (\lambda)$  and  $ P^{(k)} (0)$ are obtained then as follows 
\begin{align}
P^{(1)}(\lambda) =& 
 \frac{d}{d\lambda} P(\lambda)
 = \frac{d}{d\lambda} [ \exp(\lambda \Omega) \cdot P(0)\cdot \exp(\lambda \Omega^\dag )]  \nonumber \\
 =& \exp(\lambda \Omega) \cdot [ \Omega P(0)+ P(0) \Omega^\dag] \cdot \exp(\lambda \Omega^\dag )  \nonumber \\
 =& \exp(\lambda \Omega) \cdot  P^{(1)}(0) \cdot \exp(\lambda \Omega^\dag ) 
\label{eq:Taylor-1}
\end{align}
where  $P^{(1)}(0) = [\Omega P(0)+ P(0) \Omega^\dag]$
is identified 
as  $\left. \frac{\partial P(t)}{\partial \lambda}\right|_{\lambda=0}$,  
and 
\begin{align}
P^{(k)}(\lambda) =& \frac{d^k}{d\lambda^k} P(\lambda)   =
  \frac{d}{d\lambda} P^{(k-1)}(\lambda)  \nonumber  \\
 =& \frac{d}{d\lambda} [ \exp(\lambda \Omega) \cdot P^{(k-1)}(0)\cdot \exp(\lambda \Omega^\dag )]  \nonumber \\
 =& \exp(\lambda \Omega)  [ \Omega P^{(k-1)} (0)+ P^{(k-1)}(0) \Omega^\dag]  \exp(\lambda \Omega^\dag )  \nonumber \\
 =& \exp(\lambda \Omega) \cdot  P^{(k)}(0) \cdot \exp(\lambda \Omega^\dag ) 
\end{align}
 where 
\begin{equation}
P^{(k)}(0) = \left.\frac{\partial P^{(k-1)}(t)}{\partial \lambda}\right|_{\lambda=0}
           = [\Omega P^{(k-1)} (0)+ P^{(k-1)}(0) \Omega^\dag]. 
\label{eq:Taylor-k}
\end{equation}
When evaluating $P^{(k)}(\lambda)$ we used the fact that
 $\exp(\lambda \Omega) \Omega  =  \Omega  \exp(\lambda \Omega)$ 
(commutation)  which is consequence of the fact that both  $\Omega$ and  $\exp(\lambda \Omega)$  have the same
 eigenvectors.

Collecting Eqs. \ref{eq:Taylor-1}-\ref{eq:Taylor-k},  inserting them into Eq. \ref{eq:Taylor}, 
and setting $\lambda=1$  leads to the following  expression for the time evolution of density matrix 
\begin{align}
P(t) =&  P(0) + \{ \Omega, P^0 \} + \frac{1}{2!} \{ \Omega, \{ \Omega, P^0 \} \} + \ldots \nonumber \\
      &+ \frac{1}{k!}     \{ \Omega, \{ \ldots \{ \Omega, P^0 \} \} +\ldots
\label{eq:commutator-expansion-general}
\end{align}
where $P^0=P(0)$  and
 the curly bracket commutator  is  defined  as  $\{A,B \}= AB+BA^\dag$. 
Eq. \ref{eq:commutator-expansion-general}  can be formally written  as 
\begin{align}
P(t) =&  \exp(\hat {\mathcal{ W}}) \cdot P(0)
\label{eq:semigroup-exponentiation}
\end{align}
where the action of  the superoperator $\hat{\mathcal{ W}} = \{ \Omega,\cdot\}$   on the  density matrix $P $ is defined as
 $\hat{\mathcal{W}} P  \stackrel{\text{def}}{=}  (\Omega P + P \Omega^\dag)$, and  $\Omega = -\frac{\imath}{\hbar} \int_0^t \mathcal H(\tau)d\tau$
is a time-ordered Hamiltonian integral obtained from Magnus expansion (Eqs. \ref{eq:omega} -\ref{eq:omega3}).The results  obtained  in   Eqs. \ref{eq:semigroup-exponentiation}  and \ref{eq:commutator-expansion-general} emphasize its connection to strongly continuous one-parameter $C_0$ semigroup, \cite{PDE} which also describes the time evolution of parabolic (dissipative irreversible processes, such heat diffusion) and hyperbolic (time-reversible and energy conserving Wave dynamics) differential  equations.

The expression in  Eq. \ref{eq:commutator-expansion-general} is derived without 
constraints on the types of $P$, $H$ and $\Omega$ matrices  and is valid for arbitrary $P$ and $\Omega$. 
It 
can be used to  propagate the density   from the times $t$ to  $(t+\Delta t)$.
To  do that we truncate the
  Magnus expansion to, for example, the first order  and make use of  Eq. \ref{eq:omega1a}.   
Then $\Omega \approx  -\frac{\imath}{\hbar} \Delta t \cdot \vb{\bar{H}} $  
 where  $\vb{\bar{H}} = \frac{1}{2} \big( \vb{H}(t) + \vb{H}(t+\Delta t)\big)$
  is an average Hamiltonian matrix   between time $t$ and $(t+\Delta t)$.
 The resulting  scheme  for  density matrix  propagation is
\begin{align}
\vb{P}(t+\Delta t) =&   \vb{P}(t)  - \frac{\imath}{\hbar} \Delta t [\vb{\bar{H}, P}(t)] - 
\frac{1}{\hbar^2}  \frac{\Delta t^2}{ 2!} [\vb{\bar{H}, [\bar{H}, P}(t)] ]  \nonumber \\ 
   & +  
\frac{\imath}{\hbar^3}  \frac{\Delta t^3}{ 3!} [\vb{\bar{H}, [\bar{H}, [\bar{H}, P}(t)]]] +\ldots \nonumber \\
& + \Big( \frac{-\imath}{\hbar}\Big)^k\cdot\frac{\Delta t^k}{ k!}[\vb{\bar{H},\ldots, [\bar{H}, [\bar{H}, P}(t)]]] + \ldots
\label{eq:commutator-expansion}
\end{align} 
Employing  the commutator expansion technique enables the  density matrix propagation to  be always implemented  in terms of
matrix-matrix multiplications without a need for  diagonalization. Since diagonalization is much slower on massively parallel computers than matrix multiplication, this implementation results in substantial speedup in parallel execution.
Because the expression for $\vb{\bar{H}}$ includes $\vb{H}(t+\Delta t)$ which depends on $\vb{P}(t+\Delta t)$ itself, 
the propagation $\vb{P}(t)\to \vb{P}(t+\Delta t)$ needs to be done self-consistently  in each  iteration, 
using a more accurate  $\vb{\bar{H}}$  until  the change in  $\vb{H}(t+\Delta t)$  between iterations is negligible.
Self consistency is particularly important for stable long-time propagation (see Ref. \cite{lvnmd}).

\subsubsection{ Time evolution of the periodic Kohn-Sham state}

We now consider time evolution of a periodic Kohn-Sham state, | $\Psi_{n,{\bf k}}(x,t) \rangle$, given by 
\begin{equation}
\imath \hbar \frac{\partial{}} {\partial t} |\Psi_{n,{\bf k}}(x,t) \rangle   
=    H_{KS}(x,t) | \Psi_{n,{\bf k}}(x,t)  \rangle,
\label{eq:tdse}
\end{equation}
where $\Psi_{n,{\bf k}}(x,t)$ is the position-  and time-dependent Bloch wave function for $k$-point {\bf k} and quantum number
 $n$. It can be written as a linear combination of $N$ basis functions $\{ \phi_\alpha^\eta: \alpha=1,\ldots N\}$ 
\begin{equation}
| \Psi_{n,{\bf k}}(x,t) \rangle = \sum_{\eta=-\infty}^{+\infty}  \sum_{\alpha=1}^N   
    c_{\bf k}^{n,\alpha}(t) e^{\imath {\bf k L_\eta} } | \phi_\alpha^\eta (x) \rangle
\label{eq:psi_t}
\end{equation}
with $\eta$ describing  the $\eta$-th periodic image of the unit cell,  $L$  is a lattice constant and $\phi_\alpha^\eta(x)  = \phi_\alpha^0(x - L\cdot \eta)$.
The time-dependent Kohn-Sham Hamiltonian, $H_{KS}(x,t)$ in Eq. \ref{eq:tdse}, consists of the usual, time-independent DFT hamiltonian, $H_0[\rho(x)]$, and an external perturbation  $V(x,t)$,  which depends  on time and position 
\begin{equation}
   H_{KS}= H_0(x) + V(x,t).  
\end{equation}
The  symbol $\rho(x)$ denotes the electron density
\begin{equation}
\rho(x)= N_k^{-1}\sum_{n,k} f_{n,k} |\Psi_{n,k}|^2
\end{equation}
with occupation numbers  $f_{n,\bf{k}}$.
Inserting Eq. \ref{eq:psi_t} into \ref{eq:tdse}   and multiplying on the left  with $\langle \phi_{\alpha'}^0 | $ leads to
\begin{eqnarray}
    \imath \hbar\sum_{\alpha=1}^N \left( \sum_{\eta=-\infty}^{+\infty} e^{\imath kL\eta}\langle \phi_{\alpha'}^0 |  \phi_\alpha^\eta \rangle\right) \dot c_{\bf{k}}^{n,\alpha} (t)  = \nonumber \\ 
    \sum_{\alpha=1}^N 
    \left( \sum_\eta e^{\imath kL\eta}  \langle \phi^0_{\alpha'} | H| \phi^\eta_\alpha \rangle \right)  c_{\bf{k}}^{n,\alpha} (t)
\end{eqnarray}
where $\dot c_{\bf{k}}^{n,\alpha} (t)  = \frac{\partial}{\partial t}  c_{\bf{k}}^{n,\alpha}(t) $. The last expression can be written  as
\begin{eqnarray}
     \imath \hbar  \sum_{\alpha=1}^N  S^{\alpha',\alpha}_{\bf{k}}
\cdot \dot c_{\bf{k}}^{n,\alpha} =  
\sum_{\alpha=1}^N  H^{\alpha',\alpha}_{\bf{k}}
\cdot c_{\bf{k}}^{n,\alpha},
\end{eqnarray}
where  
\begin{eqnarray}
\label{eq:S-def}    S^{\alpha',\alpha}_{\bf{k}} =&  \sum_{\eta=-\infty}^{+\infty} e^{\imath kL\eta} \langle \phi_{\alpha'}^0 |  \phi_\alpha^\eta \rangle \\
\label{eq:H-def}    H^{\alpha',\alpha}_{\bf{k}}  = & \sum_{\eta=-\infty}^{+\infty} e^{\imath kL\eta} \langle \phi_{\alpha'}^0 | H| \phi_\alpha^\eta \rangle 
\end{eqnarray}
or in a matrix form  corresponding  to  a given $k$-point {\bf k} as 
\begin{eqnarray}
 \imath \hbar  \mathbf{S_k  \dot C_k(t) } = \mathbf{H_k  C_k(t) },
 \label{eq:tdse_nonort}
\end{eqnarray}
where ${\mathbf C_k}(t)$ is a matrix composed of column vectors $\{\mathbf C_k^1,\mathbf C_k^2,\ldots,\mathbf C_k^N\}$
 with each $\mathbf C_k^i$  vector  containing  coefficients  $c_k^{i,j}$
 of an $i$-th  Kohn-Sham state  in Eq. \ref{eq:psi_t}; ${\mathbf S_k}$ 
 and ${\mathbf H_k}$ are  overlap and hamiltonian matrices with  matrix elements  given by 
  Eqs. \ref{eq:S-def} and \ref{eq:H-def}, respectively.


Eq. \ref{eq:tdse_nonort} can be written in an equivalent Liouville-von Neumann form describing 
 density matrix propagation
\begin{eqnarray}
 \mathbf{ \dot P_k(t) } =  -\frac{\imath}{\hbar}( \mathbf{ S^{-1}_k  H_k P_k   - P_k H_k S_k^{-1}}). 
 \label{eq:LvN_nonort}
\end{eqnarray}
Here, $\mathbf{ P_k}$ is a one-particle $k$-dependent  density matrix that describes an $N_e$ electron system 
($Tr[\mathbf{P_kS_k}]=N_e$). $\mathbf{ P_k  }$ can be calculated as
 a sum of the density  contribution from  each occupied state as
 $\mathbf{ P_k  } =  \sum_{n=1}^N f_{n,k} \cdot \mathbf { P_{k,n} } $, where $f_{n,k}$  
is  an occupation number  of  the  $n$th  state  in Eq. \ref{eq:tdse_nonort},  and 
$\mathbf P_{k,n}= \mathbf{ C_{k}^{n} (C^n_k)^\dag }$ describes  its corresponding  contribution to the total density matrix,
$\mathbf{P_k}$,  with the matrix  elements given by  
$ P_{\mathbf k,n}^{\mu,\nu} =  c_{k}^{n,\mu} (\bar c^{n,\nu}_k)^\dag   $.

Eq. \ref{eq:LvN_nonort} describes  propagation of density matrix in a non-orthogonal basis set.
It can  be  obtained through the following steps: 
(a) direct differentiation of the density matrix expression,
$\mathbf{\dot P_{k,n}} = \frac{\partial}{\partial t} \left( \mathbf{ C_{k}^{n} \cdot (C^n_k)^\dag } \right)
=\mathbf { \dot{C}_{k}^{n} (C^n_k)^\dag_k + C_{k}^{n} ( \dot {C}_k^n)^\dag} $, 
(b) replacing $\vb{\dot{C}_{k}^{n}}$ and $\vb{( \dot{C}_{k}^{n}})^\dag$ with Eq. \ref{eq:tdse_nonort} 
multiplied by  $\frac{1}{\imath \hbar} \mathbf{ S}^{-1}$ on the left  and with its complex conjugation, 
(c) noting that $(\vb{S}$  and $\vb{H}$ are hermitian, thus  $(\vb{S^{-1}H)^\dag= HS^{-1}}$,  and finally
(d) summing over all occupied  states $\vb{ \dot P_k  } =  \sum_{n=1}^N f_{n,k} \cdot \vb{ \dot P_{k,n} } $.

Eq. \ref{eq:LvN_nonort} can  be transformed to an equivalent orthogonal  basis form in Eq. \ref{eq:LvN}, 
whose solution is given by  Eq. \ref{eq:commutator-expansion}. 
The  orthogonalization  can  be achieved  through  factorization of the overlap matrix $\mathbf{ S=LR}$ 
with, for example,   L\"owdin scheme ($\vb{L=R=S^{1/2}}$), or  Cholesky decomposition into 
 lower ($\vb{L}$) and  upper triangular  ($\vb{R}$)  matrices with $\mathbf{L=R^\dag}$.
Then, the  transformation between the orthogonal ($\vb{o}$ subscript) and non-orthogonal ($\vb{n}$ subscript)
basis sets for the  Hamiltonian, density, and vectors  are given by $\mathbf{H_n= L H_o R }$,
$\mathbf{H_o= L^{-1}  H_n R^{-1}}$,  $\mathbf{P_n= R^{-1} H_o L^{-1} }$,   $\mathbf{P_o= R H_n L }$,
$\mathbf{C_n= R^{-1} C_o }$, and $\mathbf{C_o= R C_n }$. See Appendix A for discussion. 

Eq.  \ref{eq:LvN_nonort} can also be rewritten in the form 
\begin{eqnarray}
 \mathbf{ \dot P_k(t) } =  \hat{\vb{ W}} \bullet \vb{P} 
 \label{eq:LvN_nonort-v2}
\end{eqnarray}
where the   superoperator  $\hat{\vb{ W}}= \{ \frac{1}{\imath \hbar}  \vb{S^{-1}H}, \cdot \}$, acts on the density matrix $\vb{P}$ 
and its result is  $\hat{\vb{ W}} \bullet \vb{P} =-\frac{\imath}{\hbar}( \mathbf{ S^{-1}_k  H_k P_k   - P_k H_k S_k^{-1}})$. 
 The  solution to Eq. \ref{eq:LvN_nonort-v2}  can be formally written as
\begin{align}
\vb{ P_k}(t) =  \exp(\mathcal{ W}) \cdot  \vb{P_k}(0),
\end{align}
 where  $\mathcal{ W}$ denotes the time-ordered integral of $\vb{\hat W}$, 
 $\mathcal{ W}= \int_0^t \vb{\hat W} d\tau=  \{\vb{\Omega},\cdot \}$, 
and Magnus expansion (Eqs. \ref{eq:omega}-\ref{eq:omega3})  is used to evaluate $\vb{\Omega}$.
The action of $\mathcal{ W}= \{\vb{\Omega}, \cdot \}$ on $\vb{P_k}$ is defined as   
 $\mathcal{ W} \bullet \vb{P_k} = \{ \vb{\Omega, P_k}  \}= (\vb{\Omega P_k +P_k\Omega^\dag})$.
For propagation of density between times $t$ and ($t+\Delta t$) with  the Magnus expansion truncated to first order 
(see Eqs. \ref{eq:omega1} and \ref{eq:omega1a}), 
the resulting  $\vb{\Omega= \Omega_1}$ and 
\begin{align}
\nonumber  
 \vb{\Omega_1} & =  -\tfrac{\imath}{\hbar} \vb{ \int_0^t S^{-1}_k  H_k(\tau) d\tau } 
   =  -\tfrac{\imath}{\hbar}   \vb{ S^{-1}_k \int_0^t H_k(\tau) d\tau }  \\
  &  = ( -\tfrac{\imath}{\hbar} \Delta t )\cdot \vb{ S^{-1}_k \bar  H},
\end{align}
where 
$\mathbf{  S^{-1}_k}$ is assumed  to not depend on time
and $\vb{\bar H}$ describes a mean  Hamiltonian,  $\vb{\bar H}=\tfrac{1}{2}(\vb{H}(t)+\vb{H}(t+\Delta t))$. 
 The final  scheme  for  the  density matrix  propagation is  
\begin{align}
\vb{P}(t+\Delta t) &=   \vb{P}^0   +  \Delta \vb{P}^{(1)} +  \Delta \vb{P}^{(2)} +  \Delta \vb{P}^{(3)} +  \ldots 
\label{eq:commutator-expansion-nonort-sum}
\end{align}
where  
\begin{align}
\nonumber       \vb{P}^0  &=  \vb{P}(t) \\
\nonumber \Delta \vb{P}^{(1)} &=  \vb{  \{ \Omega_1, P^0 \} } 
   = ( -\tfrac{\imath}{\hbar} \Delta t ) \vb{   (S^{-1}\bar{H} P^0 - P^0 \bar{H} S^{-1}) } \\
\nonumber \Delta \vb{P}^{(2)} &=  \vb{  \{ \Omega_1,\Delta  P^{(1)} \} }  \\
\nonumber    &= \tfrac{1}{2}  \left(  -\tfrac{\imath}{\hbar} \Delta t\right) \vb{   (S^{-1}\bar{H} \Delta P^{(1)} - \Delta P^{(1)} \bar{H} S^{-1}) } \\
\nonumber  \ldots& &\\
\nonumber \Delta \vb{P}^{(k)} &=  \vb{  \{ \Omega_1,\Delta  P^{(k-1)} \} }  \\
  &= \tfrac{1}{k}  \left(  -\tfrac{\imath}{\hbar} \Delta t\right) \vb{   (S^{-1}\bar{H} \Delta P^{(k-1)} - \Delta P^{(k-1)} \bar{H} S^{-1}) } 
\label{eq:commutator-expansion-nonort}
\end{align}
 For  a special case when  $\vb{S}$ is a unit matrix (for example, periodic systems 
with k-sampling limited to the $\Gamma$  point is only), 
 the matrix expression in parenthesis reduces to a  conventional commutator $[\vb{\bar H, \Delta P}^{(k-1)}]$ 
and  the  Eqs. \ref{eq:commutator-expansion-nonort} simplify to:
\begin{align}
\nonumber       \vb{P}^0  &=  \vb{P}(t) \\
\nonumber \Delta \vb{P}^{(1)} &=  \vb{  \{ \Omega_1, P^0 \} } 
   = ( -\tfrac{\imath}{\hbar} \Delta t ) \vb{   [\bar{H} , P^0 ]  } \\
\nonumber \Delta \vb{P}^{(2)} &=  \vb{  \{ \Omega_1,\Delta  P^{(1)} \} }  
   = \tfrac{1}{2}  \left(  -\tfrac{\imath}{\hbar} \Delta t\right) \vb{  [ \bar{H}, \Delta P^{(1)}]   } \\
\nonumber  \ldots& &\\
\Delta \vb{P}^{(k)} &=  \vb{  \{ \Omega_1,\Delta  P^{(k-1)} \} }  
   = \tfrac{1}{k}  \left(  -\tfrac{\imath}{\hbar} \Delta t\right) \vb{[ \bar{H}, \Delta P^{(k-1)} ]}    
\label{eq:commutator-expansion-ort}
\end{align}
which are  identical  with  Eq. \ref{eq:commutator-expansion}.
In Appendix A.2  we  show  that the expressions  in
 Eqs. \ref{eq:commutator-expansion-nonort-sum}-\ref{eq:commutator-expansion-nonort}  
can be obtained by transforming Eq. \ref{eq:LvN_nonort} to an orthogonal basis set in which  $\vb{S=1}$,  
applying the commutator expansion, Eq. \ref{eq:commutator-expansion}, 
and  finally, transforming the results back to a non-orthogonal basis set.

\section{Implementation}
\label{sec:implementation}

The implementation of real-time TDDFT presented in this paper is based on the open-source multigrid-based software package \texttt{RMG}, which can perform electronic structure calculations in real space for molecules, surfaces and periodic solids with thousands of atoms in the unit cell. 
The current implementation utilizes the so-called length gauge, which is suitable for finite systems. For periodic systems, simulations are more naturally performed in the velocity gauge, which involves perturbations expressed in terms of the current operator rather than the position operator\cite{periodic-operator} and will be pursued in future work. The velocity gauge approach  requires careful treatment of non-local potentials and pseudopotential operators.\cite{Luber-rt-tddft} 

In this section, we discuss the length-gauge implementation for molecules and large clusters, first presenting RT-TDDFT algorithm as incorporated into  \texttt{RMG}, and then  discuss optimization and parallelization.


\subsection{Algorithm}

\begin{algorithm}[h!]
\caption{Time-Dependent DFT Algorithm}
\label{alg:tddft}
\begin{algorithmic}[1] 
\scriptsize

\item[{\bf /* Ground state DFT  calculations: */ }  ]
 \State   generate $|\Psi_k\rangle$  orbitals, $H_{KS}$ Hamiltonian
 \State   select active space for TDDFT (Nocc,Nvirt)
 \State   calculate  dipole moment matrices:
 \item[]   $M_x=\langle\psi_i| x |\psi_j\rangle$;  $M_y=\langle\psi_i| y |\psi_j\rangle$;  $M_z=\langle\psi_i|z |\psi_j\rangle$
 \item[]   $\vec M = [M_x,M_y, M_z ]$
 \item[]
\item[{\bf /* Read in  parameters for RT-TDDFT: */}] 
 \State Time loop: $t_0$, $dt$, $N_{steps}$ 
 \State Convergence criteria:  
    \item[] $H_{thrs}=1\times 10^{-6}$;   $dP_{thrs}=1\times 10^{-10}$ 
 \State  External field  potential: 
    \item[]  $E_x(t)$; $E_y(t)$; $E_z(t)$

\item[] 
\item[{\bf /* Prepare initial  RT-TDDFT conditions: */}] 
 \State   Generate initial  density matrix: $P_0$
 \State   Hamiltonian and perturbation matrices:
    \item[]  $H_0\gets H_{KS}$;  $H_{-1}\gets H_0$
    \item[]  $V = E_x(t)\cdot M_x+E_y(t)\cdot M_y+ E_z(t)\cdot M_z$

\item[] 
\item[{\bf /* Initialization  of TDDFT time  loop: */}] 
\For{$j = 1$ to $N_{steps}$}
    \State $t = t_0 + j \cdot dt$
    \item[] \hspace{5pt}  {\bf /* Extrapolate $H_1$ from last  2 steps: */}
    \State          $H_1 = 2H_0 - H_{-1}$
    \item[] 
    \item[] \hspace{0pt} {\bf /*  Hamiltonian  SCF loop: */}
    \State  Init loop with dummy error: $H_{\text{err}} = 2 \cdot H_{\text{thrs}}$
    \While{($H_{\text{err}} > H_{\text{thrs}}$)}
        \item[] \hspace{15pt}  {\bf /* Magnus expansion: */ }
        \State   $\Omega = (-\tfrac{i}{\hbar} dt )\tfrac{1}{2}(H_0 + H_1 + V_{j-1} + V_{j})$

        \item[]
        \item[] \hspace{15pt}{\bf /*  Commutator expansion: */ }
        \State   $P_1 \gets  P_0$  
        \State  $dP_{\text{err}} = 2 \cdot dP_{\text{thrs}}$,
        \State   k=0 
        \While{($dP_{\text{err}} > dP_{\text{thrs}}$) }
            \State k++
            \State  $\Delta P_1 \gets  1/k \cdot (\Omega \cdot \Delta P_1 + \Delta P_1 \cdot \Omega^\dag)$
            \State $P_1 \gets  P_1 + \Delta P_1$
            \State $dP_{\text{err}} = \max(\text{abs}(\Delta P_1))$
        \EndWhile
        \item[]
        \item[]\hspace{10pt} {{\bf /*  Hamiltonians update: */  }}
        \State Convergence error: $H_{\text{err}} = \max |H_1 - H_1^{\text{old}}|$
        \State  $H_1^{\text{old}} \gets  H_1$; $H_1  \gets  DFT(\rho(P_1)  )$
    \EndWhile
    \item[]
    \item[] \hspace{0pt} {\bf /*  Analyze results: */}
    \State  print:  $\vec \mu(t) = \vec \mu_0 + \text{trace}(P \cdot \vec M)$
    \item[]
    \item[] \hspace{0pt}  {\bf /* Prepare for the next time step: */ }
    \State   $H_{-1}\gets H_0$; $H_0 \gets  H_1$; $P_0 \gets  P_1$
\EndFor
\end{algorithmic}
\end{algorithm}

An overview of the RT-TDDFT scheme   implemented  in \texttt{RMG} is shown in Algorithm~\ref{alg:tddft}.
The computational  scheme is based on  density matrix propagation 
(Eqs. \ref{eq:commutator-expansion-nonort-sum}-\ref{eq:commutator-expansion-nonort})  
and  involves following major steps:
\begin{enumerate}
\item Evaluation  of  the DFT ground state   (\texttt{lines 1-3}).
\item  Preparation of the initial conditions  and simulation  setup (\texttt{lines 4-8}).
\item  Time integration (\texttt{lines 9-29}), involving
       commutator   expansion used for  exponentiation of  the Magnus   operator, and   
       the self-consistent Hamiltonian loop.
\end{enumerate}

In the current implementation, we focused on molecular systems. Therefore, 
the DFT ground state calculations are performed at  the $\Gamma$ point. The occupied orbitals and a small number of virtual orbitals are determined self-consistently together with the ground state charge density and potentials. Extra virtual orbitals can be obtained with the non-SCF option in RMG using the fixed ground-state potential. From our tests in section 4, a large number of virtual orbitals is required to obtain a converged optical absorption spectrum. Fortunately, this number is not necessarily proportional to the system size. For example, around the same number of virtual orbitals will generate a converged spectrum for both benzene (30 electrons) and the C$_{60}$ molecule (240 electrons). 
The generated Kohn-Sham orbitals $|\psi_i\rangle$  form an orthogonal basis set for density matrix propagation. The overlap matrix in the previous section is an identity matrix.  
In general, semi-core electrons may need to be included for accurate DFT calculations, typically for heavy elements, such as the Ag nanorods in one of the examples. For some physical problems, the semi-core orbitals do not contribute to the investigated properties and can be excluded from the TDDFT propagation. In these cases, the semi-core orbitals' charge density is fixed during the TDDFT propagation. In section 4, we will show that the absorption spectra extracted from TDDFT simulations for Ag nanorods are nearly identical with or without semi-core orbitals included in the propagation.   

A subset of occupied orbitals $N_{occ}$ and a sufficient number of virtual orbitals $N_{virt}$  define the active simulation space (\texttt{line 2}) 
   with  the sizes of Hamiltonian 
and density matrices used in RT-TDDFT equal to $(N_{occ}+N_{virt})\times (N_{occ}+N_{virt})$.
Following the selection  of the active space,  the dipole operator matrices
 $M_q=\langle \psi_i|q| \psi_j \rangle$ are formed  ($q=x,y,z$) along with the initial density matrix  $P_0$,
and the Hamiltonian matrix $H_0=\langle \psi_i|H| \psi_j \rangle$. 
In the current implementation, the external perturbation  potential 
 that couples the occupied and virtual states  is defined  as the electric field dipolar matrix  (\texttt{line 8})
$\langle \psi_i|V| \psi_j \rangle$, which depends on a user-defined field strength $(E_x(t), E_y(t), E_z(t))$. In the current implementation, the constant electric field across the molecule is represented by a sawtooth-like potential, with a dipole correction in the vacuum region \cite{bengtsson-99}

The time   integration  of density matrix propagation is performed within the \texttt{lines 9-29}.
The subscript -1, 0  and 1  refer  to  time stepping and denote, respectively, matrices at times
 $(t-dt)$, $t$  and $(t+dt)$.  \texttt{Lines 15-23}   describe the recursive updates of the propagated  
density matrix  $P_1$  with the successive $\Delta P_1$ corrections to the density matrix
in Eqs. \ref{eq:commutator-expansion-nonort-sum}-\ref{eq:commutator-expansion-ort}, until convergence is  reached. 
Convergence of the commutator expansion is achieved when  $L_\infty$ norm  of $\Delta P_1$ 
is smaller than a predefined threshold, $dP_{err}< dP_{thrs}$.
The  propagation of the density matrix  from the time $t$ to $(t+dt)$ requires Hamiltonian matrices at
both times, $H_0$ and $H_1$, to evaluate the Magnus operator (\texttt{line 14}). 
However, the $H_1$ Hamiltonian matrix depends on the  propagated density $P_1$.
To account for this, the density matrix  propagation is carried out self-consistently until the difference between 
the updated $H_1$  and $H_1$ from the previous iteration (\texttt{line 25}) is smaller than the predefined threshold, $H_{thrs}$.
During the first  iteration at any given time step, the  $H_1$ is extrapolated 
from  $H_0$ and $H_{-1}$ (\texttt{line 11}).


\subsection{ Software optimization}

The real-time TDDFT module in \texttt{RMG} has been  optimized for massively parallel execution  on hybrid CPU-GPU  computing platforms.
The parallelization strategy for the main RMG routines employs a mixed programming model that utilizes MPI, OpenMP, Posix threads, and optimized libraries (OpenBlas, Scalapack, Magma-GPU). Domain decomposition is used to assign different regions of space to different cores, and a domain may span an entire node, consisting of several CPU cores and GPU accelerators. This greatly reduces the required communications and makes the communication latency more uniform. The orbitals are processed in blocks and do local synchronization within nodes at specific points of the SCF cycle, greatly reducing the need for global synchronization. The RMG code scales nearly linearly to hundreds or more CPU-GPU nodes, depending on problem size, and utilizes all CPU cores and all GPUs in each node.

The numerical operations required to implement the propagator described in 
Eqs.  \ref{eq:commutator-expansion-nonort-sum}-\ref{eq:commutator-expansion-ort} 
 are dominated by matrix-matrix operations in $[\vb{H}, \Delta \vb{P}^{(k)}]$, 
which exhibit $O(N^3)$  scaling as the number of basis functions $N=(N_{occ}+N_{virt})$ increases. 
(The basis functions refer to Kohn-Sham orbitals in this case). While these are computationally intensive, highly 
optimized BLAS libraries are usually available for performing them on both CPU and GPU architectures. 
Therefore, the key to achieving good performance is structuring the code to leverage these libraries and,
in the case of systems equipped with GPU accelerators, minimizing data transfers between CPU and GPU memory spaces.

The domain decomposition algorithm employed by RMG means that objects of interest, such as the Kohn-Sham 
wavefunctions, ionic potentials, and charge density are mapped to three-dimensional regions of real space. 
The domains are distributed over MPI tasks via a Cartesian mesh. Computing resources assigned to each task 
include a number of host CPU cores per task as well as GPU accelerators for systems equipped with them. 
Depending on the specific system configuration, GPUs may be shared with several tasks running on the same 
compute node. 
This paradigm means that each task has a copy of every basis function but only for a specific region of real space,  
making computation of matrix elements, such as $\langle\psi_i|H|\psi_j \rangle$, straightforward, although a reduction 
operation is required to combine the partial results from all MPI tasks.  Copies of the resultant 
square overlap matrices of order $N$ are also needed by all MPI tasks to perform rotations of the orbitals. 
The same is true of the matrices $\vb{P}$, $\Delta P$, $\vb{\bar H}$ from 
Eqs. \ref{eq:commutator-expansion-nonort-sum}-\ref{eq:commutator-expansion-ort}, 
which presents an interesting optimization problem.  

First, while the commutator expansion (Eqs. \ref{eq:commutator-expansion-nonort-sum}-\ref{eq:commutator-expansion-ort})
implemented  in RMG avoids  diagonalization, the resulting density matrix $\vb{P_1}$  
is  arbitrarily close to the results obtained with exact diagonalization propagator 
(subject to the convergence threshold  $dP_{thrs}$). 
The computational cost  of the scheme  is  determined  by  the cost of this commutator,
 $\vb{[\bar H, \Delta P]}$. Thus, the optimization effort focuses on its minimization.
The  matrices $\vb{\bar H}$ and  $\vb{\Delta P}$  used  in RT-TDDFT are usually  hermitian 
and can be decomposed  into a sum  of symmetric and antisymmetric matrices holding, respectively, the real and imaginary parts,  
$\vb{\bar H = S +\imath A}$ and $\vb{\Delta P= S' + \imath  A'}$. 
Then, the complex-complex commutator can be implemented as  a sum  of four separate real-real  commutators
\begin{equation}
\vb{ [\bar H, \Delta P] = [S,S'] - [A,A'] + \imath[S,A'] +\imath[A,S']},
\label{eq:commutator}
\end{equation}
each having the same computational cost. However, for pure DFT functionals using only the  $\Gamma$ point, 
the Hamiltonian  matrix $\vb{ \bar H}$ is purely real and symmetric $\vb{S}$, with the imaginary 
part $\vb{A}$ vanishing. This reduces the computational cost by 50\% over 
the full complex-complex commutator, since the commutator expression becomes
$\vb{ [\bar H, \Delta P] = [S,S'] + \imath[S,A']}$.
Next, both commutators can be  implemented via two  matrix multiplications, or alternatively as a single 
matrix multiplication followed by a transpose
\begin{align}
\nonumber \vb{[S,S']} =& \vb{SS' -S'S } =\vb{  SS' - (SS')}^T \\
\nonumber \vb{[A,S']} =& \vb{AS' -S'A } =\vb{  AS' + (AS')}^T 
\end{align}
The computational 
cost of the matrix multiplications in commutator Eq. \ref{eq:commutator} increases as $O(N^3)$, 
while the size of the matrix and hence communication costs between nodes only increase as $O(N^2)$. 
For small $N$ it is 
faster for each MPI task to duplicate the work in Eq. \ref{eq:commutator}. 
As $N$ increases, there will be a point where 
a distributed matrix multiplication will be faster. The exact value of the crossover depends on 
both $N$ and the relative speed of communication to computation on a given hardware platform. On ORNL's Frontier, 
using a node-local multi-GPU matrix multiply routine is faster than a single GPU once $N$ exceeds a few thousands. 
For a large enough $N$, a fully distributed matrix multiplication routine will be the best solution, 
but we have not yet performed calculations in that size range. 
Eq. \ref{eq:commutator} may be formulated using 
only matrix multiplies or a combination of matrix multiplications and transposes. Theoretically, a transpose 
scales as $O(N^2)$, but in practice, the memory access patterns required by a transpose are so inefficient 
that it is not a clear win unless suitably optimized routines are available. 
Both AMD and Nvidia provide such routines via the \texttt{DGEAM} and \texttt{ZGEAM}  
functions, so the commutator was implemented in that form.

\section{Numerical Results}
\label{sec:results}

In this section, we benchmark our current RT-TDDFT implementation across molecular systems of varying sizes, focusing on practical aspects: (1) convergence of the results with respect to the number of virtual orbitals, (2) time step choice, (3) conservation of energy and long-term stability, and (4) applicability to large molecular systems. The simulations aim to answer key questions: How does our real-time TDDFT implementation compare with others? What is the stability of time-integration over long time simulations? What is the maximum practical time step and system size for simulations? Addressing these questions helps to understand its limitations and provides a baseline for its applicability.

The benchmarking covers a range of molecular systems: organic (benzene), inorganic (fullerene), bioorganic (chlorophylls), and metallic (silver nanoparticles). Benzene, a benchmark molecule for real-time TDDFT, is used to compare the stability of our time integration scheme and to perform the time-step stress test. Chlorophylls are used for long simulation length tests, while silver nanoparticles are used for system size tests.

We simulate the optical response to a linearly polarized electric "kick" impulse (Dirac delta) perturbation
\begin{equation}
V^{i,j}(t) =   \delta(t) \sum_{q}^{x,y,z}  E_q\cdot \langle \phi_i| \hat \mu_q | \phi_j \rangle
\end{equation} 
where  $ V^{i,j}(t)$ is a matrix element of external  electric field  perturbation (length gauge) over  the  set of active orbitals (occupied and virtual), $q$ runs over $\{x,y,z\}$ cartesian coordinates,    $\delta(t)$ is  a Dirac delta time envelope (kick-type perturbation), and  $\hat \mu_q= q$ is a dipole moment operator. 
Photoabsorption spectra are calculated as  a sum of Fourier transforms of  $\mu_x(t)$,  $\mu_y(t)$   and  $\mu_z(t)$  obtained from  three independent simulations with the kick perturbation potential polarized along $x$, $y$, and $z$  directions, respectively.
 The current implementation is validated against popular real-time TDDFT implementations in \texttt{NWChem} and \texttt{Q-Chem} (localized Gaussian basis sets), and against linear-response TDDFT in \texttt{Quantum ESPRESSO} (\texttt{QE}). Both \texttt{RMG} and \texttt{QE} use similar lattice and mesh setups with Vanderbilt's norm-conserving pseudopotentials \cite{Vanderbilt}.
All calculations were performed with the semi-local PBE functional \cite{pbe}  due to its computational efficiency. While hybrid functionals, which incorporate a fraction of exact exchange, can provide more accurate descriptions of electronic excitations, their computational cost in real-time TDDFT is significantly higher,  with the exchange operator becoming complex-valued \cite{lvnmd}.

\subsection{Small and medium  systems}

In this subsection, we discuss the quality of agreement between the absorption spectra for benzene and C$_{60}$ fullerene generated using the RMG time propagation-based TD-DFT implementation and implementations found in a popular open-source a plane wave pseudopotential DFT code (quantum ESPRESSO, QE)\cite{Vanderbilt,QE-TDDFT2,QE} and an all electron TD-DFT implementation relying on Gaussian-type orbitals as basis functions (\texttt{NWChem})\cite{lopata-2011}. All TD-DFT calculations carried out with periodic-boundary conditions (PBC) codes (\texttt{RMG} and \texttt{QE}) rely on a large (>10 a$_0$) vacuum region surrounding the isolated systems to prevent spurious interactions with the periodic images.

\begin{figure}[ht!]
    \centering
    \includegraphics[width=\columnwidth]{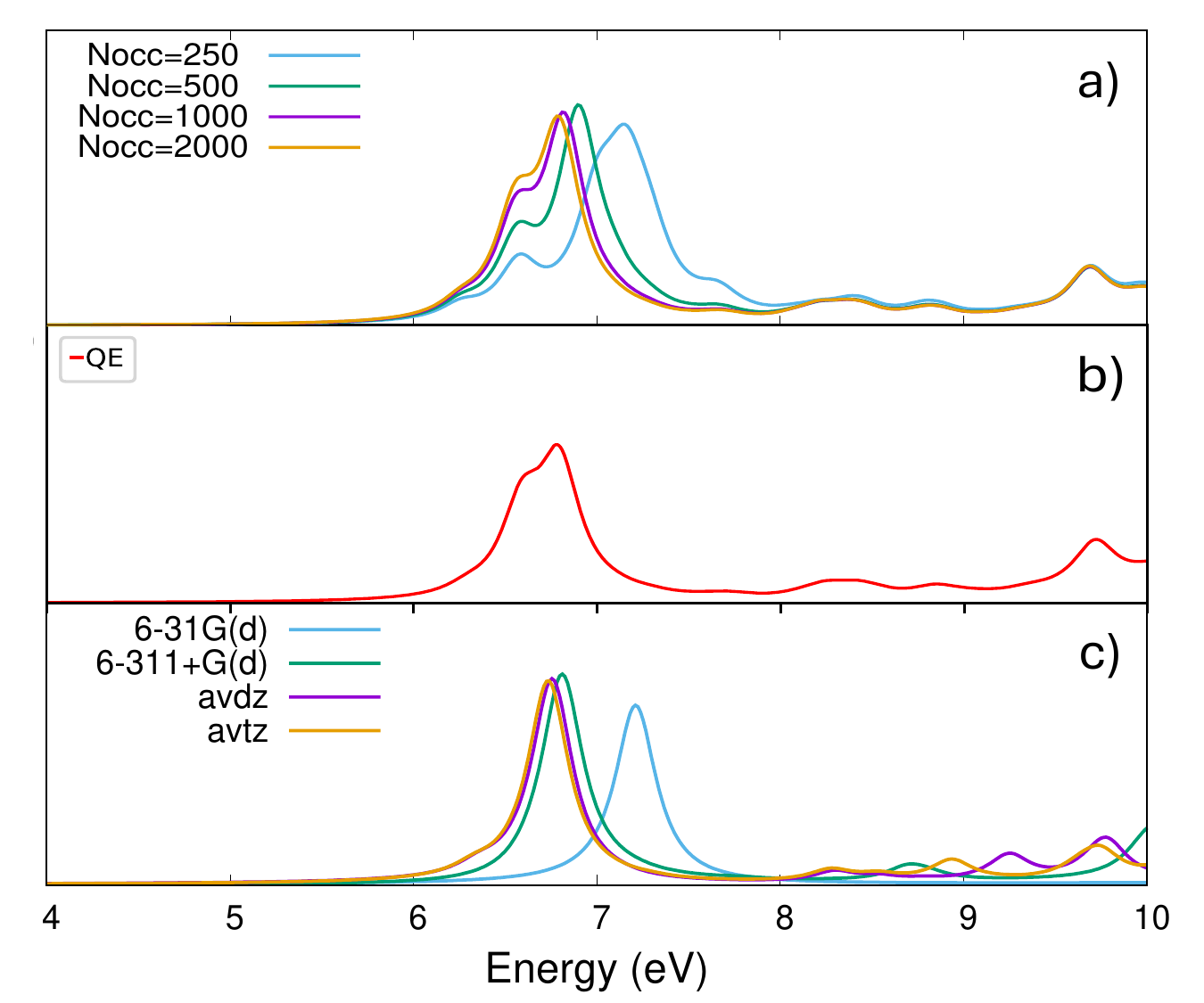}
    \caption{\footnotesize  Comparison of benzene dipole oscillator strength (sub-ionization energy region) across the different TD-DFT simulations platforms, with (a)  the RMG results using different number of unoccupied orbitals in the expansion space, (b) the quantum ESPRESSO result using analogous plane-wave cutoff to the RMG grid, and well-converged with respect to the depth of the Lanczos recursion employed, and (c) the NWChem result using a typical basis set for TD-DFT calculations of systems comprised of main group elements.  }
    \label{fig:benzene}
\end{figure}

\begin{figure}[ht!]
    \centering
    \includegraphics[width=\columnwidth]{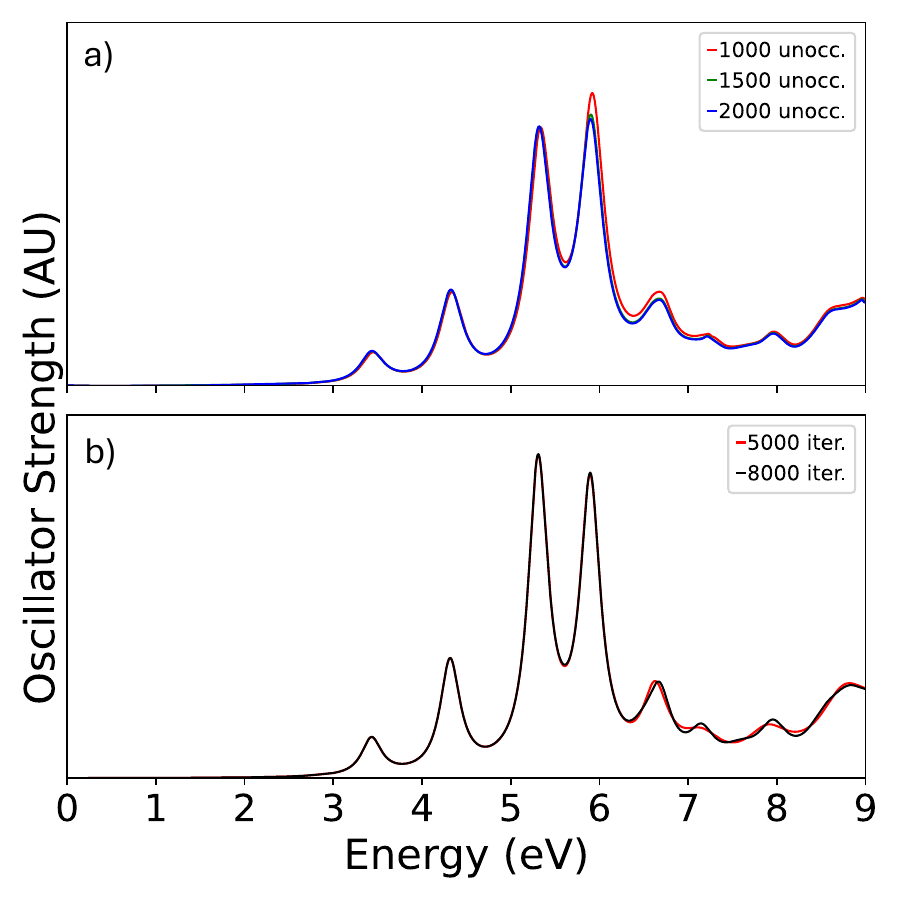}
    \caption{\footnotesize  Comparison of C$_{60}$ oscillator strength showing the impact of (a) increasing the number of unoccupied states utilized in the RMG TD-DFT propagation scheme, compared with (b) increasing the number of iterations of the Lanczos recursion-based solver for the analogous system in the quantum ESPRESSO TD-DFT implementation.} 
    \label{fig:c60}
\end{figure}

Figures  \ref{fig:benzene}  and \ref{fig:c60}  show absorption spectra for  benzene and C$_{60}$ buckminster fullerene  for the energy  range below the ionization threshold. For \texttt{NWChem} we used  two Pople  and two Dunning basis sets:   6-31G(d),  6-311+G(d),  aug-cc-pVDZ  and aug-cc-pVTZ.  For  the benzene molecule, a lattice of 36  $\times$ 36 $\times$ 30 ($a_0$) is used with periodic boundary conditions.  We used  120$\times$ 120 $\times$ 100 real space grid for RMG, and 25 Ry wave function cutoff energy with  120$\times$ 120 $\times$ 96 FFT grid for QE.  For the C$_{60}$ fullerene, a 42 $\times$ 42 $\times$ 42 ($a_0$) lattice is used. We employed a 120$\times$ 120 $\times$ 120 real space grid for RMG,  and  30 Ry wave function cutoff energy with  150$\times$ 150 $\times$ 150 FFT grid for QE. 
Excellent agreement is observed between the line shapes of the photoabsorption spectra produced by the \texttt{RMG} and \texttt{QE} codes using  the same boundary conditions and delocalized basis sets. At the same time, a good qualitative agreement is also observed with the all electron \texttt{NWChem} results, where the use of Gaussian-type orbitals imposes the asymptotic decay of the electron density to zero (see fig. \ref{fig:benzene}).

For benzene, the experimentally  observed photoabsorption  spectrum is mainly comprised of three $\pi\to \pi^*$ transitions: a very weak band  at 4.90 eV   ($^1A_{1g}\to {}^1B_{2u}$), a stronger (vibronic) band at 6.19 eV  ($^1A_{1g}\to {}^1B_{1u}$), and a very intense band  at 6.96  eV ($^1A_{1g}\to {}^1E_{1u}$) \cite{benzenene-experiment:2002}.
In our simulations, the absorption peak for the ($^1A_{1g}\to {}^1E_{1u}$) transition is observed  in the range  between   6.9 eV  and 7.2 eV, depending on the number of basis functions (NWChem) or the  number of unoccupied orbitals (RMG) used. The vibronic transition is forbidden by symmetry and does not show up unless the molecular structure of benzene is distorted  from $D_{6h}$  symmetry as a result of nuclear motion.

\begin{figure*}[ht!]
    \centering
    \includegraphics[width=0.99\textwidth]{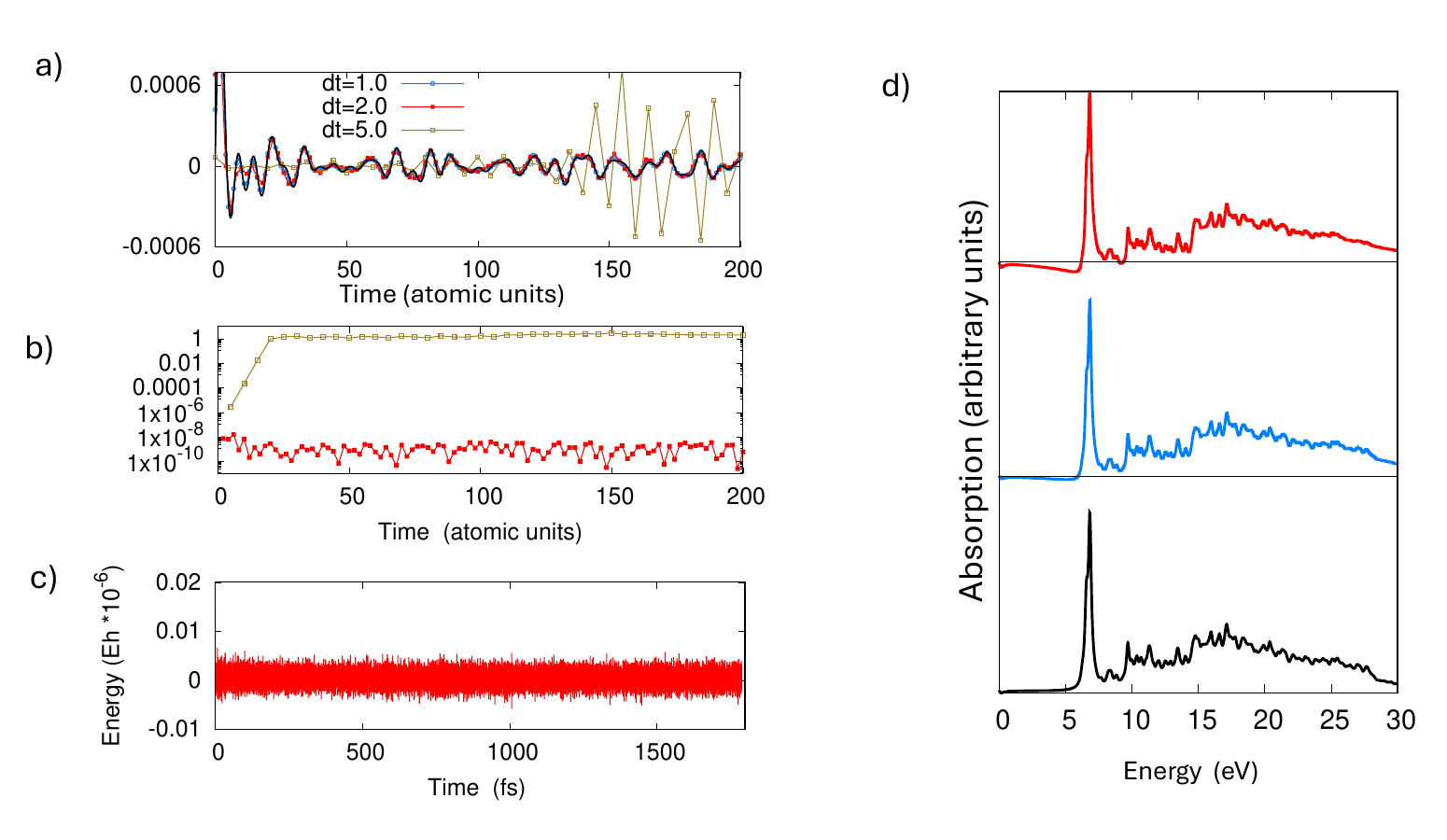}
    \caption{\footnotesize  
    RT-TDDFT results for benzene, showing time-step sensitivity and energy conservation. (a) Time dependence of dipole moment for time steps of  0.2 (solid black line), 1.0, 2.0, and 5.0 a.u.t., with instability and scaled oscillations for dt = 5.0 a.u.t. (b)  Comparison of energy conservation  (logarithmic  scale, energy in $E_h$)  up to 200 a.u.t. for dt = 2.0 and 5.0 a.u.t., highlighting energy loss at larger time steps. (c) Long time  energy conservation for dt = 2.0 a.u.t. over 1.8 ps. (d) Photoabsorption spectra for dt = 0.2, 1.0, and 2.0 a.u.t., with dt = 0.2 a.u.t. as the reference.}
    \label{fig:benzene-dt}
\end{figure*}

Increasing the  number of basis functions in simulations consistently decreases the excitation energy for the $^1A_{1g}\to {}^1E_{1u}$ transition, in accordance with the variational principle applied to the first  excited state. 
Result from NWChem simulations with the smallest 6-31G(d) basis set  correspond to  RMG results with 250 unoccupied orbitals, both showing the photoabsorption peak  near 7.2 eV.  The position of the $^1A_{1g}\to {}^1E_{1u}$ transition  converges at about ~6.9  eV with 1000 unoccupied RMG  orbitals,  which corresponds to similar NWChem results obtained with aug-cc-pVDZ basis set. 
Similar convergence with respect to the number of unoccupied orbitals has been  observed for RMG simulations of C$_{60}$ fullerene, as shown  in Figure \ref{fig:c60}.   For the fewest number of unoccupied orbitals used in RMG (1000), the convergences is nearly complete even at higher energies. Meanwhile, the low-energy region of the spectrum is well converged within the QE implementation with relatively few iterations of the Lanczos algorithm, with higher energy peaks converging much more slowly.  Both spectra show a reasonable agreement to the solution-phase experimental spectra \cite{c60-experiment}.     
Thus, in the following RMG simulations  we  use 1000 unoccupied orbitals as a baseline for  RT-TDDFT modeling.

Notably, the QE and RMG results for benzene show a  lower intensity   absorption peak  at  6.6 eV  accompanying  the  main   $^1A_{1g}\to ^1E_{1u}$ transition at 6.9 eV.  This peak corresponds to  $^1A_{1g}\to {}^1B_{2u}$  transition  and  is an artifact of  PBE functional \cite{Rocha-benzene:2024}. A similar peak, albeit less pronounced, has also been observed in NWChem simulations  using a localized basis set. An in-depth analysis of theoretical benzene spectra and comparison between  localized Gaussians and grid based results is provided  elsewhere \cite{Rocha-benzene:2024}.

Next, we analyzed the sensitivity of the photoabsorption spectra to the size of the time step  and the long-term stability of the time-integration algorithm implemented  in RMG.
The time steps used in the simulations ranged from 0.2 a.u.t. (atomic unit of time)  to  5 a.u.t.  
Figure  \ref{fig:benzene-dt} shows the time-dependence of the instantaneous dipole moment for benzene,  conservation of energy, and  comparison  of the resulting  photoabsorption spectra.  The simulated  photoabsorption spectra for  time steps up to dt=2 a.u.t. show very good agreement with reference simulations using a time step of 0.2 a.u.t., as seen  in  Figure \ref{fig:benzene-dt}(d).
Clearly, the time integration algorithm is very stable for up to 2 a.u.t., demonstrated by  good agreement between time-dependent dipole moments and reference results (black solid line) obtained with a time step of 0.2 a.u.t., as seen in Figure \ref{fig:benzene-dt}(a).   
 For  a time step 5.0 a.u.t.,  the time-integration  becomes unstable, with the dipole moment oscillating wildly over time (see olive line in \ref{fig:benzene-dt}.(a)) and loss of energy conservation  within the first few  time steps (see \ref{fig:benzene-dt}(b)). 
Figure \ref{fig:benzene-dt}(c) shows conservation of energy for simulations with a time step  of 2.0 a.u.t for  a 1.8 picoseconds simulation, demonstrating excellent conservation of energy and no observable energy drift throughout the  entire simulation  time.

Finally, we compared our time integration in \texttt{RMG} with a similar Magnus expansion propagator implemented in \texttt{NWChem} and with the Modified Midpoint Unitary Transform (MMUT) (implemented in NWChem and in Q-Chem)\cite{lopata-2011,rt-tddft-qchem}.
Comparing the time-propagation in \texttt{NWChem} and \texttt{RMG}, we observed a very small energy drift on the order of 10$^{-6}$ $E_h$ in \texttt{NWChem} simulations, whereas \texttt{RMG} did not show any noticeable energy drift. It is noteworthy that both \texttt{NWChem} and \texttt{RMG} propagators are based on the time-integration algorithm of density matrices originally proposed by Jakowski and Morokuma \cite{lvnmd}. The main difference in the implementations is that \texttt{NWChem} uses a fixed number of Hamiltonian matrix updates (predictor/corrector iterations), namely two, whereas time propagation in \texttt{RMG} and in Ref. [\citenum{lvnmd}] relies on dynamic convergence testing measured as the $L_\infty$ norm of Hamiltonian matrix updates. In \texttt{RMG}, we used a Hamiltonian matrix updates threshold set to 10$^{-7}$ a.u.

As expected, increasing the simulation time step in RMG requires an increased number of Hamiltonian update iterations. For RMG simulations of benzene with time steps of 0.2, 0.5, 1.0, and 2.0 a.u.t., the observed number of Hamiltonian update iterations per time step in the initial non-equilibrium simulation phase were 3, 4, 5, and 9, respectively. In the later simulation phase, the observed number of Hamiltonian updates dropped to 1, 2, 3, and 6 for the time steps of 0.2, 0.5, 1.0, and 2.0 a.u.t., respectively. 
These results indicate that for RT-TDDFT simulations of benzene, it is computationally beneficial to increase the time step up to 1.0 a.u.t. However, the computational cost savings are diminished when increasing the time step to 2.0 a.u.t due to the increased number of Hamiltonian updates required.


\begin{figure*}[h!]
    \centering
     \includegraphics[width=0.99\textwidth]{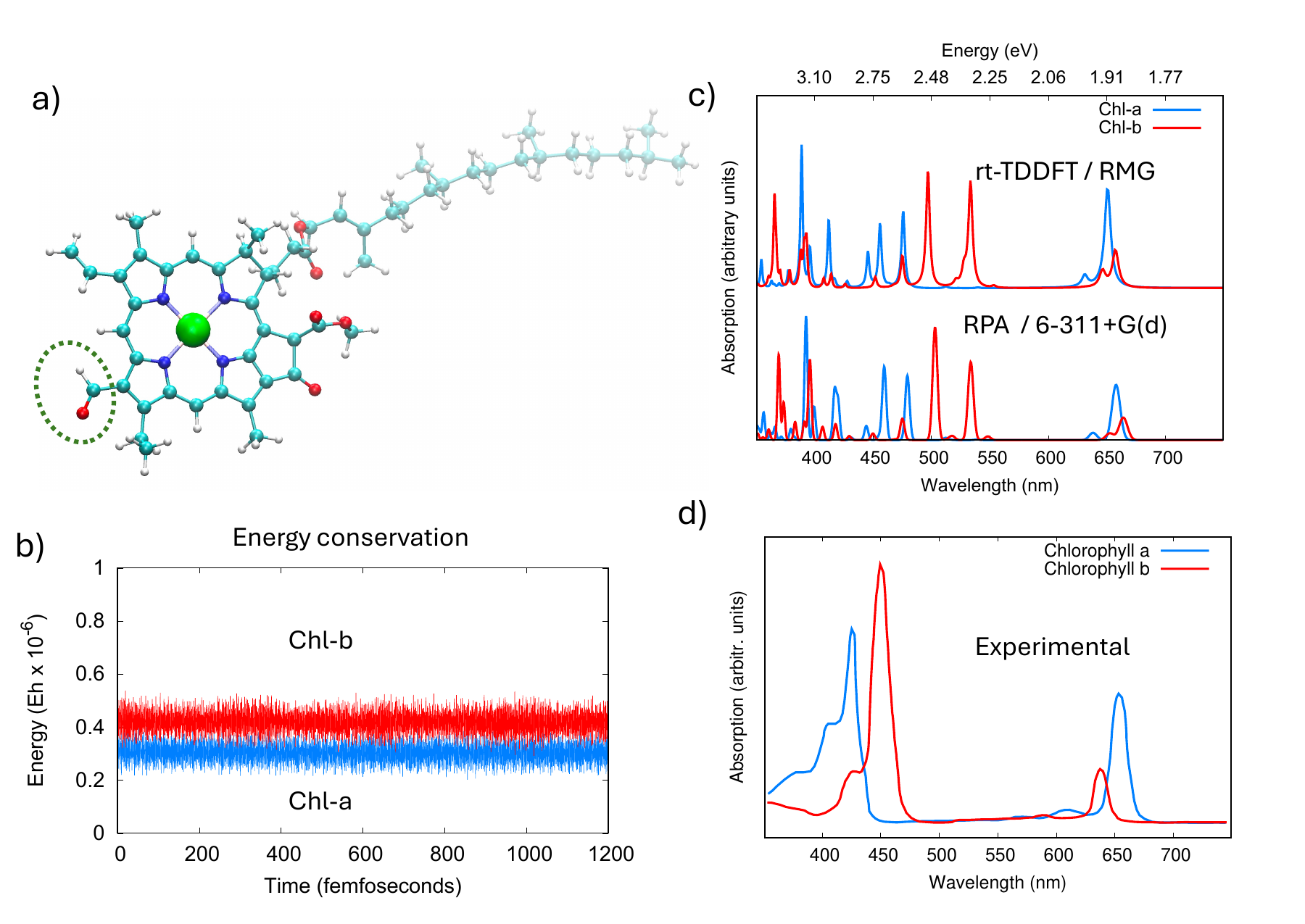}   
    \caption{\footnotesize   RT-TDDFT simulations of chlorophylls  A  nd B. (a) Molecular structure of chlorophyll B used in the current calculations, with the -CHO group (highlighted within the green dotted oval) that is substituted by a -CH$_3$ group in chlorophyll A. (b) Energy conservation for both chlorophylls, calculated relative to the first step after the perturbation potential is turned off. (c) Comparison of photoabsorption spectra for chlorophylls A and B from RMG calculations (denoted as rt-TDDFT/RMG) with linear response  TDDFT  within Random Phase Approximation (RPA) and with 6-311+G(d) basis set.The RPA results are plotted using Gaussian broadening to mimic experimental line shapes.  (d) Experimental photoabsorption spectra of chlorophylls A and B adapted from  
    Ref.[\citenum{chlorphylls}].}
    \label{fig:chlorophylls}
\end{figure*}

\subsection {Chlorophylls}

The molecular systems discussed in the previous section (benzene and C$_{60}$ fullerene) represent spatially compact molecules with high symmetry (point groups $D_{6h}$ and $I_h$, respectively). Here, we illustrate the application of our RT-TDDFT module implemented in RMG to photoactive chlorophyll molecules, which are asymmetric and significantly larger than benzene and C$_{60}$ fullerene, focusing on long-term stability  and comparison with  experimental data.

Chlorophylls are bioorganic molecules found in green plants and algae that play a critical role in photosynthesis and the conversion of CO$_2$ to carbohydrates. Several variations of chlorophylls exist in photosynthetic organisms. Chlorophyll A (C$_{55}$H$_{72}$MgN$_4$O$_5$) is the most abundant of the  chlorophylls and is directly involved in converting solar radiation into chemical energy. Chlorophyll B (C$_{55}$H$_{70}$MgN$_4$O$_6$) plays an auxiliary role in photosynthesis. Both chlorophylls contain a rigid aromatic porphyrin ring with Mg in the center and a floppy hydrocarbon tail.  Figure~\ref{fig:chlorophylls}(a)  shows the molecular structure of chlorophyll B used in the current simulations. The difference between the molecular structures of chlorophyll A and chlorophyll B is that the -CHO group (highlighted within the green dotted oval) is substituted with an -CH$_3$ group.

The molecular structures of both chlorophylls used for TDDFT simulations were obtained from structure optimization with  PBE/6-311+G(d) theory.
The RMG input files were generated using RMG web interface \cite{RMG-DFT-gui} with grid spacing of 0.3 $a_0$ and  10 Angstrom vacuum around the chlorophyll molecules to prevent self-interaction with periodic images. The resulting orthorhombic lattice was   (85.9 $a_0) \times$  (43.45  $a_0) \times$  (30.78  $a_0)$ for chlorophyll A and (86.3  $a_0) \times$   (43.3  $a_0) \times$  (32.0 $a_0)$  for chlorophyll B.
 The corresponding  values  of real-space grids were 288 $\times$ 152 $\times$ 104 and 288$\times$ 144 $\times$ 112, respectively for chlorophyll A  and chlorophyll B.
Similarly to benzene and C$_{60}$, periodic boundary conditions were employed.
The DFT simulations were performed  with the PBE functional and Vanderbilt's norm-conserving pseudopotentials \cite{Vanderbilt}.

To verify the long-term stability of our  time propagation in RMG, we performed 50,000 steps of TDDFT simulations for chlorophyll A and B with a time step of 1.0 a.u.t. and a total simulation time of 1.2 ps. The resulting conservation of energy, optical absorption spectra for both chlorophylls, and  comparison with experimental results, are shown in Figure~\ref{fig:chlorophylls}(b)-(d). The  simulations show excellent conservation of energy with small energy oscillations on the level of 10$^{-7}$ E$_h$ and no observable energy drift throughout the 1.2 ps long simulations. The time integration with a time step of 1.0 a.u.t. required about six Hamiltonian matrix updates per simulation time step during the initial, non-equilibrium phase of electron dynamics.  It was reduced to about four Hamiltonian  updates during the later stages of time propagation. For comparison purposes, we also checked the number of Hamiltonian updates for a simulation with a  reference time step of 0.2 a.u.t. The observed Hamiltonian matrix updates were reduced to only two. These results are consistent with those observed for benzene and C$_{60}$ discussed in the previous section.

Figure~\ref{fig:chlorophylls}(d)  compares RMG spectra from RT-TDDFT  (denoted  as rt-TDDFT/RMG) with  LR-TDDFT spectra within the Random Phase Approximation (RPA) and obtained withi the localized Gaussian  6-311+G(d) basis set (denoted as RPA/6-311+G(d). The agreement between both approaches is excellent. Both result in the same peaks positions  showing only some  intensity differences between both approaches.
The experimental spectra, shown in  Figure~\ref{fig:chlorophylls}(d), are from Ref. [\citenum{chlorphylls}] and were obtained for chlorophylls dissolved in diethyl ether. They exhibit absorption bands in the blue range (Soret band) with maxima near 428 nm (chl-a) and 453 nm (chl-b), and in the red optical range (Q band) with maxima near 661 nm (chl-a) and 642 nm (chl-b). The simulated spectra show very good agreement with the Q band peak maximum near 640 nm. Similarly to the experimental data, the intensity of the simulated chl-a peak is significantly stronger than that of chl-b. In the blue range, the experimental peaks are clearly composed of several overlapping peaks, which are visible in the simulated spectra with several peaks within the 350-450 nm range for chl-a and 450-520 nm for chl-b, whereas the corresponding experimental band maxima peak near 428 nm and 453 nm, respectively. We note that the TDDFT simulations are calculated in vacuum and for a single structure of chlorophyll, and as such, neglect thermal motion and broadening. TDDFT simulations by others \cite{chlorphylls-rubio} show similar agreement with experimental results, with TDDFT excitation (CAM-B3LYP) energies underestimating experimental Q-band by 0.11 eV (ca. 30 nm) for chl-a and 0.15 eV (ca. 40 nm) for chl-b.

\subsection{Plasmonic nanoparticles}

Simulations discussed in the previous sections focused on comparing the photoabsorption spectra obtained by RMG with other codes for small and medium-size molecules and benchmarking the effect of  time step on the long-term stability of the time integration. Here, we test the feasibility of our TDDFT implementation for simulating electron dynamics in large molecular systems.  Specifically, we address the following questions: What is the largest molecular system that can be modeled with the current TDDFT implementation? What are the limiting factors? What is the computational cost per time step for the largest feasible system?

\begin{table*}[h!]
    \centering \footnotesize
    \begin{tabular}{l|r|r|r}
    \hline \hline
System                            &  Ag$_{540}$ (4.9nm)  & Ag$_{1620}$  (14.7nm)  & Ag$_{2160}$ (19.6nm)   \\ 
Number of active electrons        &     5,940     &  17,820        &  23,760        \\

Number of basis functions (total)  &   5630          &  16,890       &  24,520          \\
Number of basis functions (active) &   3470          &  10,410       &  15,880          \\
Number of Frontiers nodes used    &    16         &  32            & 64             \\ \hline
Hupdate                           &   32.58 (sec) &  153.42 (sec)  & 270.35 (sec)    \\
Rho                               &   31.48 (sec) &  124.21 (sec)  & 162.00 (sec)   \\
Vh                                &   13.08 (sec) &   11.50 (sec)  & 8.74 (sec)   \\
ELDYN                             &    0.86 (sec) &   56.37 (sec)  & 129.40 (sec)   \\
exchange/correlation              &   13.40 (sec) &   11.39 (sec)  & 9.28 (sec)     \\
Total TD-DFT time                 &  110.06 (sec) &  378.6 (sec)   & 634.3 (sec)    \\    
\hline
Time for a single TDDFT time step &    1.1 (sec)  &  3.78 (sec)    & 6.34 (sec)    \\
Time for a single  SCF iteration  & 27.0          &   124 (sec)    & 108 (sec)     \\ 
\hline\hline 
    \end{tabular}
    \caption{ \footnotesize  Explicit real-time TDDFT timings  on  Frontier supercomputer for 100 TDDFT time steps for selected three Ag-nanorods: Ag$_{540}$, Ag$_{1620}$ and  Ag$_{2160}$. The number of active basis functions corresponds to the size of  density matrix propagated spanning \emph{4d,5s} valence orbitals  (11 electrons per each Ag atom)  plus a  set of unoccupied orbitals. The total number of basis function is larger and includes  semicore orbitals in addition to valence ones (19 electrons per Ag atom).}
    \label{tab:Frontier-timing}
\end{table*}

As a model system, we used silver nanorods of increasing size to simulate the longitudinal plasmon resonance peak. Metal nanoparticles have garnered significant attention for their applications in nanoplasmonics and nanophotonics, offering a wide range of applications in sensing, energy, medicine, and imaging \cite{PhysToday-nanoplasmonics,OpticsExpress-nanoplasmonics}. Nanoplasmonics bridges the gap between nanoscale materials science and optical physics, enabling applications in optical filters, waveguides, photonic-circuit components, and sensors \cite{plasmonic-resonators,wong-plasmonics-waveguide-2018,wong-plasmonic-antennas-2017}.

\begin{figure*}[h!]
    \centering
    \includegraphics[width=0.99\textwidth]{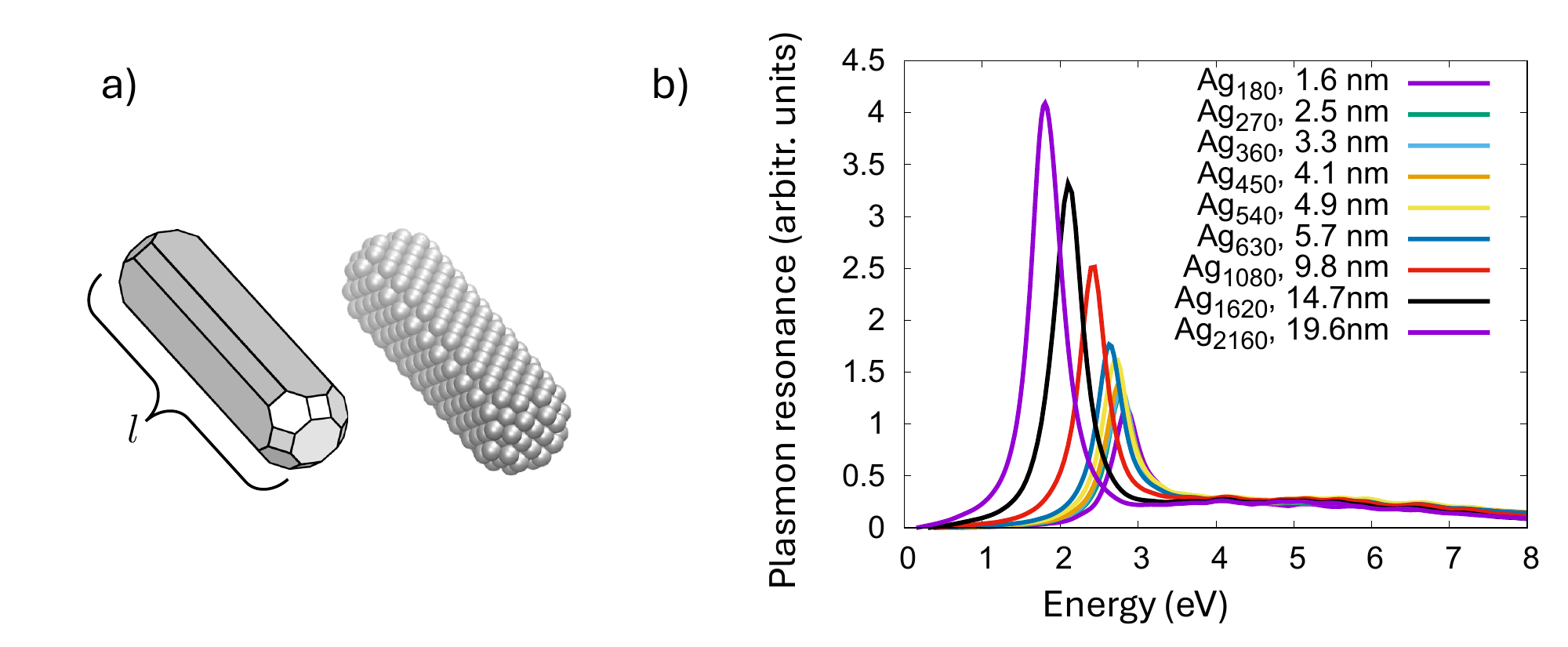}
    \caption{\footnotesize  Comparison of longitudinal (i.e., the component  along the long axis) dipole oscillator strength normalized with respect to the number of Ag atoms across different lengths ($l$) of elongated  rhombicuboctahedral silver nanorods, showing the expected increase in intensity and redshift in the dipolar longitudinal localized surface plasmon resonance with increasing system size.}
    \label{fig:ag_rods}
\end{figure*}
Nanoplasmonic phenomena typically emerge for metallic nanoparticles ranging in size from 2 to 20 nm \cite{OpticsExpress-nanoplasmonics}. For optimal performance, metal nanoparticles for nanoplasmonic applications should be smaller than approximately 25 nm (the metal's skin depth), to allow penetration by optical radiation, and larger than about 2 nm, which corresponds to the distance an electron with Fermi velocity travels during a single light oscillation. This size range is significantly smaller than the 400 nm wavelength of visible light \cite{PhysToday-nanoplasmonics}. Utilizing such narrow metal films enables surface-enhanced spectroscopy (SES) applications, such as surface-enhanced Raman spectroscopy (SERS).

A silver nanorod exhibits two orthogonal plasmon resonances: a long-wavelength longitudinal plasmon resonance for light polarized along its long axis, and a shorter-wavelength transverse plasmon resonance for light polarized perpendicular to its long axis.  Here, we investigate  a longitudinal surface  plasmon resonance (SPR), focusing on the shift of the SPR  frequency as a function of the nanorod size.   We used a sequence of   increasing size  rhombicuboctahedral  silver nanorods  of the dimensions  $l$  $\times$ (1.2 nm) $\times$ (1.2 nm),   where $l$  ranges  from 1.6 nm  to  20 nm.  
For all silver  nanoparticle cases, the TDDFT active space  included the 11  highest energy electrons from each Ag atom, excluding the semicore electrons (frozen core approximation) and 1000-2000 unoccupied (virtual) orbitals.
Overall, we considered  nine   silver nanorods:  Ag$_{180}$, Ag$_{270}$,  Ag$_{360}$, Ag$_{450}$, Ag$_{540}$,  Ag$_{630}$, Ag$_{1080}$, Ag$_{1620}$, and  Ag$_{2160}$  (see Figure~\ref{fig:ag_rods}),  with the corresponding number of electrons considered  in the TDDFT being  1980,  2970,  3960,  4950,  5940,  6930,  11,880,   17,820,  and  23,760, respectively.  The RMG calculations used an orthorhombic  lattice with the dimensions $L_x$ $\times$ (44 $a_0$)   $\times$ (44 $a_0$), and $L_x = \{$52, 68, 84, 100, 114, 130, 208, 300, 392$\}$. The corresponding numbers of real space grid points were $N_x \times 132 \times 132$ with $N_x = \{$156, 204,  252,  300, 342,  390,  624, 900, 1176$\}$.

All TDDFT calculations for silver nanorods were performed on the Frontier supercomputer at the Oak Ridge Leadership Computing Facility using up to 64 nodes and utilizing all GPUs (8 per Frontier node). The resulting TDDFT spectra obtained with 1.0 a.u.t time step, showing the longitudinal plasmon resonance peak for all nanorods are displayed in Figure~\ref{fig:ag_rods}(b). As expected, we observe a red shift in the position of the plasmon resonance peak with increasing nanoparticle size, with a nearly linear dependence between the particle size and the wavelength corresponding to the plasmon resonance \cite{plasmon-size-dependence-2022}.

Table~\ref{tab:Frontier-timing} shows the explicit timings for real-time TDDFT simulations on Frontier for three Ag-nanorods: Ag$_{540}$,  Ag$_{1620}$ and Ag$_{2160}$. Notice that broken sub timing shown for Ag$_{540}$, Ag$_{1620}$ and  Ag$_{2160}$ does not add up to the total of 110.06 sec, 378.6  sec and 634.3 sec, respectively. The reason is that unless one adds in specific synchronization calls, GPU kernel launches are asynchronous with respect to the host. The synchronization points can cause an overall drop in performance and prevent overlapping computation with data transfer. 

For a given system, both the TDDFT and SCF portion of the calculations exhibit O(N$^3$) scaling per step, but the prefactor for TDDFT steps is smaller. This can be readily seen for the Ag$_{2160}$ system run on 64 nodes of Frontier (using all of its CPUs and GPUs) where a TDDFT step requires approximately 6.34  seconds while an SCF step requires 108 seconds.  The   TDDFT simulations for the next size  Ag nanorod, namely Ag$_{2700}$,  should be feasible with the current code but have not yet been performed. 
Calculations on even larger systems present no conceptual problem but may require further optimization of GPU memory usage in order to maintain a similar level of performance.

\section{Summary and Outlook}

In this work, we presented theory,  implementation and benchmarking of a real-time TDDFT (RT-TDDFT) module developed within the RMG code. This approach is designed to model the electronic response of molecular systems to external perturbations in real-time, providing insights into non-equilibrium dynamics and excited states. Our benchmarking spanned a wide range of molecular systems, from small organic molecules like benzene to large metallic silver nanorods.

The benchmarking demonstrated excellent agreement of RMG results with other TDDFT implementations (e.g., NWChem and Quantum ESPRESSO), validating the accuracy and robustness of our approach. Furthermore, our time-integration algorithm showed superior stability, allowing for long-term simulations with minimal energy drift, even for large time steps.

In addition to electronic dynamics, future extensions of our RT-TDDFT implementation will incorporate nuclear motion and spin dynamics, enabling a comprehensive description of coupled electron-nuclear and spin processes. The inclusion of nuclear motion will allow for the study of non-adiabatic dynamics, which are essential for understanding photoinduced processes and energy transfer in molecular systems. Spin dynamics will facilitate the investigation of magnetic properties and spintronics applications, where the interplay between spin and charge degrees of freedom is crucial.

This implementation  opens new avenues for exploring electronic dynamics in complex systems, which are   computationally challenging for traditional methods. The scalability and efficiency of the RMG code on massively parallel architectures, such as the Frontier supercomputer, allow to tackle large-scale problems, including simulations of plasmonic nanoparticles of  thousands of atoms. Future work will focus on optimizing the current implementation to push the limits of system size and simulation time further, and exploring more sophisticated electronic structure methods beyond DFT.

Excellent performance and large scale capabilities of RT-TDDFT within RMG  provide  valuable tools for the study of photoactive materials, nanoscale devices, and other systems where electronic dynamics play a crucial role. The planned extensions include nuclear and spin dynamics, providing  insights into the fundamental processes governing complex molecular and material systems.

\section{Acknowledgements}

This research was conducted at the Center for Nanophase Materials Sciences (CNMS), 
which is a US Department of Energy, Office of Science User Facility at Oak Ridge National Laboratory.
This research used resources of the Oak Ridge Leadership Computing Facility at the Oak Ridge National Laboratory, 
which is supported by the Office of Science of the U.S. Department of Energy under Contract No. DE-AC05-00OR22725. The development of RMG for exascale computers was supported by DOE's Exascale Computing Project (ECP), Project Number: 17-SC-20-SC, for DFT input to QMCPACK. Computing resources were provided through  the Innovative and Novel Computational Impact on Theory and Experiment (INCITE) program
and computational resources of the ACCESS (Advanced Cyberinfrastructure Coordination Ecosystem: Services \& Support) program through allocation TG-DMR110037. 

\appendix

\section{Appendix} 
\subsection{Orthogonalization} 
{ \bf Rules for Transformation  between non-orthogonal and orthonormalized basis  set representations.}

The non-diagonal overlap matrix  $\mathbf { S_k}$ in Eqs. \ref{eq:tdse_nonort} and \ref{eq:LvN_nonort} represents a  Gram matrix of inner products 
 over non-orthogonal basis set vectors $\{  \phi_i^n \}$, which is positive definite  and
 its  matrix elements are
 $\langle \phi_i^n|  \phi_j^n  \rangle = S^{i,j}$, 
The  superscript $n$ denotes the non-orthogonalility of the basis set. 
It is often more convenient to work with an equivalent orthonormalized  basis set,  
$\{  \phi_i^o \}$,  such that  $\langle \phi_i^o|  \phi_j^o  \rangle =\delta_{i,j}$.
The transformation  from a non-orthogonal  basis set   $\{  \phi_i^n \}$ 
 to an  orthonormal one,    $\{  \phi_i^o \}$,  
 can be achieved by  factorization of the overlap matrix $\mathbf{ S}=\mathbf{LR}$,
 where $\mathbf L$ and $\mathbf R$  can be obtained in many different ways,
 for example, using  Cholesky or L\"owdin factorization, or by directly constructing $\mathbf L$ and $\mathbf R$ 
from the eigenvectors of $\mathbf S$. 
It can be easily checked that the inverse of $\mathbf S$ is  given by $\mathbf S^{-1}= \mathbf R^{-1} \mathbf L^{-1}$. 
Here, the  subscript $\mathbf k$ is dropped for simplicity. 

Considering two orthogonal Bloch states,  $|\Psi i\rangle$  and  $|\Psi j\rangle$,  with the corresponding coefficients 
in a non-orthogonal basis set $\mathbf C_n^i $ and  $\mathbf C_n^j $,  we can write
\begin{align}
\langle \Psi_i |\Psi_j \rangle =  ( {\mathbf C}_n^i)^\dag  \cdot   {\mathbf S} \cdot  {\mathbf C}_n^j  =\delta_{i,j}
\end{align}
where $ {\mathbf A}^\dag$  denotes the hermitian conjugate of  $\mathbf A$.
Inserting  the factorized overlap matrix into the last expression and using 
$(\mathbf{AB})^\dag = \mathbf{B}^\dag \cdot \mathbf{A}^\dag$ gives
\begin{align}
 ({\mathbf C}_n^i  ) ^\dag \cdot   {\mathbf S} \cdot  {\mathbf C}_n^j    
\nonumber  =& ({\mathbf C}_n^i)^\dag  ( \mathbf{L L}^{-1})   \cdot   \mathbf{S} \cdot  (\mathbf {R^{-1} R}) {\mathbf C}_n^j  \\
\nonumber  =&  ( ({\mathbf C}_n^i)^\dag  \mathbf L) \cdot  (\mathbf{L^{-1} S R^{-1}  }) \cdot (\mathbf{ R C}_n^j )   \\
\nonumber  =&  (\mathbf L^\dag {\mathbf C}_n^i)^\dag \cdot \mathbf I  \cdot  (\mathbf{ R C}_n^j )    \\
 =&    ({\mathbf C}_o^i)^\dag  \cdot \mathbf I  \cdot  \mathbf{ C}_o^j 
=\delta_{i,j}
\end{align}
Here,  $\mathbf L^\dag = \mathbf R$  (or equivalently,  $\mathbf R^\dag = \mathbf L$),
  the diagonal matrix  $\mathbf I = \mathbf{L^{-1} S R^{-1}  }$ represents a Gram matrix of the inner product
 transformed to an orthonormal basis set, and  $ \mathbf{ C}_o^j = \mathbf{ R C}_n^j$  describes 
the $|\Psi_j \rangle$ vector in orthogonalized basis representation.
In fact,  inner product of a vector is a special case of a bilinear  form  and the corresponding transformation   
between the orthonormal and non-orthogonal basis sets is also valid for Hamiltonian matrix, 
$\mathbf H_o = \mathbf{L^{-1} H_n R^{-1}  }$, where $\mathbf H_o$ and $\mathbf H_n$ denote, respectively,
the Hamiltonian matrices in orthonormal and non-orthogonal basis sets.  
Similarly,  the inverse transformation from orthogonal to non-orthogonal basis sets  are
 $ \mathbf{ C}_n = \mathbf{ R^{-1} C}_o$, and  $ \mathbf{ C}^\dag_n = \mathbf C_o^\dag  \mathbf L^{-1} $
 and  $\mathbf H_n = \mathbf{L H_o R  }$.

Following the definition,  the  density matrix in a non-orthogonal basis is   
\begin{align}
\nonumber
\mathbf P_n  =&  \mathbf{ C}_n \cdot \mathbf{ C}_n^\dag=  
( \mathbf{R^{-1} C}_o ) \cdot ( \mathbf{ C}_o^\dag  \mathbf{L^{-1} }) \\
\nonumber  =&     
 \mathbf{R^{-1}} (\mathbf{ C}_o  \cdot  \mathbf{ C}_o^\dag ) \mathbf{L^{-1} }=  \\
  =&  \mathbf{R^{-1}}  \mathbf{ P}_o  \mathbf{L^{-1}} 
\end{align}
where 
 $\mathbf P_o =  \mathbf{ C}_o \cdot \mathbf{ C}_o^\dag$  is the density matrix in an orthonormal basis set  and  the
transformation between the orthogonal and non-orthogonal density  matrices is 
$\mathbf P_n  =  \mathbf{R^{-1}}  \mathbf{ P}_o  \mathbf{L^{-1}}$ and
$\mathbf P_o  =  \mathbf{R}  \mathbf{ P}_n  \mathbf{L}$.
This can be verified by  testing  whether  the trace of the product of the density matrix and of given operator, which is a physical observable,  is invariant  of the  basis set
transformation.  For example, number of electrons is given  by 
\begin{align}
\nonumber
N_{el} =& Tr[ \mathbf{P_n S}]   \\
\nonumber    =&   Tr[ ( \mathbf{R^{-1}}  \mathbf{ P}_o  \mathbf{L^{-1}} )\cdot(\mathbf{L R} )  ]  \\
\nonumber    =&   Tr[ \mathbf{R^{-1}}   \mathbf{ P}_o (  \mathbf{L^{-1} L } )\cdot\mathbf{ R}  ]   \\
\nonumber    =&   Tr[   \mathbf{R^{-1}}   \mathbf{ P}_o   \cdot \mathbf{ I} \cdot \mathbf{ R}  ]   \\
\nonumber    =&   Tr[   \mathbf{ P}_o   \cdot   \mathbf{ I } \cdot ( \mathbf{R \cdot R^{-1}}) ]  \\
    =&  Tr[   \mathbf{ P}_o   \cdot   \mathbf{ I } ]    
\end{align}
where  the following cyclic permutation  property of trace has been used 
$Tr[\mathbf{ABC}] = Tr[\mathbf{BCA}]= Tr[\mathbf{CAB}]$.
It can be similarly verified  that $Tr[ \mathbf{P_n H_n}] = Tr[ \mathbf{P_o H_o}]$.

Finally, we can  verify  that  the  Liouville-von Neumann  equation   for the density matrix propagation  
in an orthogonal basis set, 
$\imath \hbar {\mathbf{ \dot P_o}} (t) = [\mathbf{ H_o, P_o}]$, 
transforms  to   Eq. \ref{eq:LvN_nonort} 
in a non-orthogonal basis set representation.
We start by noticing that the transformation matrices $\mathbf{ R, L, R^{-1}}$, and $\mathbf L^{-1}$
 do not depend on time, therefore
\begin{align}
\nonumber
\imath \hbar {\mathbf{ \dot P_o}} (t) &=  \imath \hbar  \frac{\partial}{\partial t} ( \mathbf{  RP_n L})
  =\imath \hbar  \mathbf{ R \dot P_n L}   \\
\nonumber  = & [\mathbf{ H_o, P_o}] = \mathbf{ H_o P_o}  - \mathbf{ P_o H_o}  \\
\nonumber  =&  \mathbf{ (L^{-1} H_n R^{-1} )( R P_n L ) }  - \mathbf{ (R P_n L)( L^{-1} H_n R^{-1}) } \\
\nonumber  =&  \mathbf{ L^{-1} H_n (R^{-1}\cdot R) P_n L  }  - \mathbf{ R P_n (L\cdot L^{-1}) H_n R^{-1} } \\
  =&  \mathbf{ L^{-1} H_n  P_n L  }  - \mathbf{ R P_n  H_n R^{-1} } 
\end{align}
Since in the above expression  $\imath \hbar {\mathbf{ \dot P_o}} (t)= \imath \hbar  \mathbf{ R \dot P_n L}$, 
 to obtain   Eq.  \ref{eq:LvN_nonort},    the last expression needs to multiplied  on the left 
by $\mathbf{ R^{-1}}$  and  on the right  by $\mathbf{ L^{-1}}$, resulting in
\begin{align}
\nonumber
\imath \hbar {\mathbf{ \dot P_n}}  
&= \mathbf{ R^{-1}} \cdot (\mathbf{ L^{-1} H_n P_n L} - \mathbf{ R P_n H_n R^{-1}}) \cdot \mathbf{L^{-1}} \\
\nonumber =& \mathbf{( R^{-1} L^{-1} ) H_n P_n (L \cdot L^{-1})}  \\
\nonumber  & - \mathbf{(R^{-1} R) P_n H_n (R^{-1} L^{-1}) }     \\
=&  \mathbf{ S^{-1}  H_n P_n   - P_n H_n   S^{-1} }
\label{eq:LvN_nonort-Appendix}
\end{align}
which is identical to  Eq.  \ref{eq:LvN_nonort}.

\subsection{ Propagation  in non-orthogonal  basis set }

 Here we show the derivation of density matrix  propagation in a non-orthogonal  basis set.
We start  from the expression   in an orthogonal basis and  apply  transformation rules  to
derive the expression in a non-orthogonal basis set.

Consider   propagation of the  density matrix in a non-orthogonal  basis set governed  by Eq. \ref{eq:LvN_nonort-Appendix} which is the same as Eq. \ref{eq:LvN_nonort}.
For simplicity we denote the non-orthogonal   density  and Hamiltonian matrices as $\mathbf{P_n}$ and  $\mathbf{H_n}$.
These  can be transformed to those in   an orthogonal basis $\mathbf{P_o}$ 
and $\mathbf{H_o}$  by a factorization of the overlap matrix $\mathbf{S= LR}$.  The expressions for 
transformations between the basis sets are  given by  
$\mathbf{H_n= L H_o R }$,  
$\mathbf{H_o= L^{-1}  H_n R^{-1}}$, and  $\mathbf{P_n= R^{-1} P_o L^{-1} }$,   and $\mathbf{P_o= R P_n L }$, and
$\mathbf{C_n= R^{-1} C_o }$, and $\mathbf{C_o= R C_n }$.

Denoting   $\gamma^k= \frac{1}{k} ( -\frac{\imath}{\hbar}\Delta t )$ and using 
$\vb{S^{-1}=  R^{-1}  L^{-1}}$, 
the time evolution for   density matrix propagation  in an orthogonal basis  set given by
\begin{align}
\nonumber \mathbf { P_o} (t+\Delta t) =&  \vb{ P_o^0}  +  \Delta \vb{P_o^{(1)}} 
+  \Delta \vb{P_o^{(2)}}   \ldots + \Delta \vb{P_o^{(k+1)}} +\ldots  \\
\nonumber   =&  \mathbf{ P_o^0}  +\gamma^1 [\vb{  H_o, P_o^0}] +   \gamma^2 [\vb{  H_o}, \Delta \vb{P_o^{(1)}}]  \ldots \\
            & + \gamma^{(k+1)}  [\vb{  H_o}, \Delta \vb{P_o^{(k)}}] 
\end{align}
and 
\begin{align}
\nonumber \vb{ P_n} & (t+\Delta t) = \vb{R^{-1} \cdot  P_o} (t) \cdot \vb{ L^{-1} } \\
\nonumber & =\vb{R^{-1}}  (\vb{ P_o^0}  +  \Delta \vb{P_o^{(1)}}  
            +  \Delta \vb{P_o^{(2)}}   \ldots + \Delta \vb{P_o^{(k)}} +\ldots   )  \vb{ L^{-1} } \\
  & =  \vb{ P_n^0}  +  \Delta \vb{P_n^{(1)}} +  \Delta \vb{P_n^{(2)}}   \ldots + \Delta \vb{P_n^{(k)}} +\ldots 
\end{align}
where
\begin{align*}
\vb{ P_o^0} = &\vb{ P_o}(t) \\
\vb{ P_n^0} = & \vb{R^{-1}} \cdot \vb{ P_o}(t) \cdot \vb{ L^{-1} }, 
\end{align*}
\begin{align*}
\Delta \vb{ P_o^{(1)}} = & \gamma^1  \vb{[H_o, P_o^0 ] } \\
                =  \gamma^1 & ( \vb{H_o P_o^0 - P_o^0 H_o  } ) \\
                =  \gamma^1 & \Big(\vb{ (L^{-1} H_n R^{-1})( R  P_n^0 L)} - \vb{(R P_n^0 L)(L^{-1} H_n R^{-1}) }\Big)\\
                =  \gamma^1 & \Big( \vb{ L^{-1} \cdot   H_n P_n^0  \cdot  L } -  \vb{ R \cdot  P_n^0 H_n\cdot R^{-1} }\Big),
\end{align*}
\begin{align*}        
\Delta \vb{ P_n^{(1)}} = & \vb{R^{-1}} \cdot \Delta \vb{ P_o^{(1)}} \cdot \vb{ L^{-1} } \\
= \gamma^1  & \vb{R^{-1}} 
    \Big(  \vb{ L^{-1} \cdot   H_n P_n^0  \cdot  L } -  \vb{ R \cdot  P_n^0 H_n\cdot R^{-1} } \Big) \vb{ L^{-1} } \\
= \gamma^1 &
    \Big( \vb{R^{-1}} \vb{ L^{-1} H_n P_n^0 L} \vb{ L^{-1} } - \vb{ \vb{R^{-1}} R  P_n^0 H_n  R^{-1}\vb{ L^{-1}}} \Big) \\
= \gamma^1 &
    \Big( \vb{S^{-1} \cdot H_n P_n^0 }  - \vb{   P_n^0 H_n\cdot S^{-1}}   \Big), 
\end{align*}
  the k-$th$ order correction
\begin{align*}
\Delta \vb{ P_o^{(k)}} = & \gamma^k  [\vb{H_o}, \Delta \vb{P_o^{(k-1)} }] \\
        =  \gamma^{k} & ( \vb{H_o}\Delta \vb{ P_o^{(k-1)}} - \Delta \vb{P_o^{(k-1)} H_o  } ) \\
        =  \gamma^{k} & \Big(\vb{ (L^{-1} H_n R^{-1})( R} \Delta \vb{P_n^{(k-1)} L)}  \\
         & {\qquad \qquad \qquad }  -  (\vb{R} \Delta \vb{ P_n^{(k-1)} L)(L^{-1} H_n R^{-1}) }\Big)\\
        =  \gamma^k & \Big( \vb{ L^{-1} \cdot   H_n } \Delta \vb{ P_n^{(k-1)}  \cdot  L } 
            -  \vb{ R} \cdot \Delta \vb{ P_n^{(k-1)} H_n\cdot R^{-1} }\Big),
\end{align*}
and 
\begin{align}        
 \Delta \vb{ P_n^{(k)}} = & \vb{R^{-1}} \cdot \Delta \vb{ P_o^{(k)}} \cdot \vb{ L^{-1} } \\
\nonumber = \gamma^{k}  & \vb{R^{-1}} 
    \Big(  \vb{ L^{-1}}  \vb{H_n} \Delta \vb{P_n^{(k-1)}  L} - \vb{ R}  \Delta \vb{P_n^{(k-1)} H_n  R^{-1} } \Big) \vb{ L^{-1} } \\
\nonumber = \gamma^{k} &
    \Big( \vb{R^{-1} L^{-1} H_n} \Delta \vb{P_n^{(k-1)} L  L^{-1} } \\
\nonumber     & {\qquad \qquad \qquad \qquad } -  \vb{R^{-1} R }\Delta \vb{P_n^{(k-1)} H_n  R^{-1} L^{-1}} \Big) \\
\nonumber = \gamma^{k} &
    \Big( \vb{S^{-1} \cdot H_n} \Delta \vb{P_n^{(k-1)} }  - \Delta \vb{ P_n^{(k-1)} H_n\cdot S^{-1}}   \Big) 
\end{align}
The results obtained here are identical with  Eqs. \ref{eq:commutator-expansion-nonort}  from direct exponentiation of  $\exp(\mathcal{W})$,  where $\mathcal{W}=\{\vb{\Omega,.} \}$.

\balance
\bibliography{references}
\balance 

\end{document}